\documentclass[final,3p,times]{elsarticle}
\usepackage{amssymb}
\usepackage{epsfig}
\usepackage{amsmath}
\usepackage{multirow}
\usepackage{natbib}
\setcitestyle{numbers,square}

\begin{document}
\begin{frontmatter}

\title{Pseudo Transitions in the Finite-Size Blume-Capel Model}

\author[label1]{Wei Liu \corref{cor1}}
\ead{weiliu@xust.edu.cn}
\author[label1]{Lei Shi}
\author[label1]{Xin Zhang}
\author[label2]{Xiang Li \corref{cor1}}
\ead{lixiang11@alumni.nudt.edu.cn}
\author[label3,label4,label5]{Fangfang Wang}
\author[label6]{Kai Qi \corref{cor1}}
\ead{kqi@mail.sim.ac.cn}
\author[label3,label4,label5]{Zengru Di}
\cortext[cor1]{Corresponding author}

\affiliation[label1]{organization={College of Sciences, Xi'an University of Science and Technology},
    city={Xi'an},
    postcode={710054}, 
    country={China}}  

\affiliation[label2]{organization={National Innovation Institute of Defense Technology},
    city={Beijing},
    postcode={100071}, 
    country={China}}  

\affiliation[label3]{organization={School of Systems Science,Beijing Normal University},
    city={Beijing},
    postcode={100875}, 
    country={China}}
    
\affiliation[label4]{organization={International Academic Center of Complex Systems, Beijing Normal University}, 
    city={Zhuhai}, 
    postcode={519087}, 
    country={China}}

\affiliation[label5]{organization={Department of Systems Science, Faculty of Arts and Sciences, Beijing Normal University},
    city={Zhuhai}, 
    postcode={519087}, 
    country={China}}

\affiliation[label6]{organization={2020 X-Lab, Shanghai Institute of Microsystem and Information Technology, Chinese Academy of Sciences},
    city={Shanghai},
    postcode={200050},
    country={China}}

\begin{abstract}
This article investigates the pseudo transitions of the Blume-Capel model on two-dimensional finite-size lattices. By employing the Wang-Landau sampling method and microcanonical inflection point analysis, we identified the positions of phase transitions as well as higher-order phase transitions. Through Metropolis sampling and canonical ensemble analysis, we determined the geometric characteristics of the system at these transition points. When the crystal field parameter $D$ exceeds 1.965, crossing the tricritical point, no third-order dependent phase transition is observed. However, a fourth-order independent transition was identified in the high-temperature region, and through Metropolis sampling analysis, a phase transition from the ordered paramagnetic phase to the disordered paramagnetic phase was confirmed, enhancing the phase diagram. Furthermore, the positions of the third-order phase transition obtained from both microcanonical and canonical analyses are consistent and mutually corroborative. We speculate that third-order dependent transitions vanish in the presence of strong first-order phase transitions.
\end{abstract}
\end{frontmatter}

\section{Introduction}
Phase transitions are ubiquitous in nature and represent a fundamental topic in classical statistical physics and thermodynamics. They are among the most striking collective phenomena in many-body systems \cite{gross2001microcanonical} and can also be interpreted as abrupt changes in the topology of the N-body configuration space \cite{franzosi1999topological, casetti2000geometric}. These sudden transitions can have catastrophic consequences, such as irreversible climate change \cite{lenton2008tipping, fan2018climate} or ecological collapse \cite{scheffer2003catastrophic, arumugam2024early}. Consequently, the precise identification of critical points and the development of early warning systems have become key research priorities. Studies have shown that systems exhibit distinct behaviors in both temporal \cite{donangelo2010early, brovkin2021past, dmitriev2023twitter} and spatial dimensions \cite{dakos2010spatial, rietkerk2021evasion, bian2024early} as they approach critical transitions. Furthermore, early warning signals have been explored through approaches such as thermodynamic quantities \cite{yan2023thermodynamic} and machine learning-based methods \cite{bury2021deep}. In this study, we aim to identify characteristic indicators of early warning signals for phase transitions in spin systems, providing a foundation for critical early-warning methodologies and potential real-world applications.

In recent developments, microcanonical analysis has increasingly been utilized for identifying phase transitions \cite{gross2001microcanonical, junghans2006microcanonical}. This method was further generalized by Qi and Bachmann \cite{qi2018classification} to identify higher-order transitions, where independent and dependent transitions were distinguished. Independent transitions are similar to traditional phase transitions, occurring independently of other processes within the system. Dependent transitions rely on the occurrence of lower-order transitions. Independent and dependent transitions occur on either side of the phase transition point, indicating potential for early warning signals of phase transitions. By applying this method, researchers have demonstrated their effectiveness in the Ising \cite{sitarachu2022evidence}, Potts \cite{wang2024exploring}, and Baxter-Wu models \cite{liu2022pseudo}, including the detection of signals indicative of third-order transitions. The use of these analytical tools has revealed early signs of phase transitions, offering a strong basis for predicting critical shifts in spin systems. A recent study on ice premelting by Tian et al \cite{hong2024imaging} has attracted significant attention. Ice premelting occurs when water molecules migrate from beneath the ice surface to the surface, disrupting the initially ordered structure. This disturbance is analogous to the way isolated spins perturb the ordered state in spin systems. The discovery of this phenomenon demonstrated the existence of third-order transitions in real-world systems, thereby further validating the theoretical framework. This discovery not only deepened the understanding of third-order transitions, but also introduced new perspectives for future research. Building on these foundations, we aim to explore phase transition behavior through the Blume-Capel model.

The Blume-Capel model, originally proposed independently by Blume \cite{blume1966theory} and Capel \cite{capel1966possibility}, has become a foundational framework for studying tricritical behavior \cite{saul1974tricritical, gunaratnam2024existence, kwak2015first, goldman1973tricritical, moueddene2024critical, azhari2020tricritical, blote2019revisiting, mandal2016geometrical}. One of its distinguishing features is its capacity to model the transition from second-order to first-order phase transitions, thereby revealing intricate critical behaviors. Researchers have investigated this model using diverse approaches, including mean-field theory \cite{blume1966theory, capel1966possibility}, renormalization group techniques \cite{antenucci2014critical}, and Monte Carlo simulations \cite{mahan1978blume, clusel2008alternative, silva2006wang}. Research on this model has covered a wide range of materials, including metallic alloys \cite{dias2011site, pena2009blume, blatter1985reversible, sinkler1997neutron}, magnetic thin films \cite{mazzitello2015far}, and superconducting thin films \cite{goldman1973tricritical}. In addition to its applications in physics, the model has been employed in sociology, particularly in recent studies on depolarization \cite{kaufman2024social, diep2024monte}. Given its critical importance and its capacity to capture a wide range of phase behaviors, the Blume-Capel model has generated numerous publications \cite{mozolenko2024blume, cirillo2024homogeneous, akin2024investigation, ovchinnikov2024influence}. This model was selected because it enables the simultaneous observation of the relationships between second-order and first-order phase transitions, each with the behaviors of third-order independent and dependent transitions.

Exploring the changes in higher-order phase transitions across a system's transition from second-order to first-order phases is an intriguing problem. Specifically, understanding how the system's geometric characteristics evolve during higher-order phase transitions is crucial for devising effective early warning strategies in complex systems, including artificial swarms \cite{bechinger2016active} and active bacterial colonies \cite{copeland2009bacterial, qi2022emergence}. The density of states (DOS) of the Blume-Capel model was obtained using the efficient Wang-Landau algorithm \cite{wang2001efficient, landau2004new, li2007numerical, wust2011unraveling, vogel2013generic, vogel2014scalable, ferreira2018wang}, while the geometric properties of the Blume-Capel model were examined through Metropolis sampling. Using these data, we examined phase transition behavior through both microcanonical and canonical methods, identifying the locations of higher-order transitions. In the second section, the model and analytical methods are introduced. The third section presents and discusses the results, and the final section provides a summary and outlines future research directions.

\section{The model and the method}
\subsection{Blume-Capel Model}

The spin-1 Blume-Capel model is a generalization of the Ising model, with the Hamiltonian defined on a two-dimensional $L \times L$ lattice as follows:
\begin{equation}
\mathcal{H} = -J \sum_{\langle ij \rangle} \sigma_i \sigma_j + D \sum_i \sigma_i^2,
\end{equation}
where \(\sigma_i\) represents the spin located on the two-dimensional square lattice, and \(\langle i, j \rangle\) denotes the sum over nearest-neighbor sites. \(\sigma_i\) can take the values -1, 0, or +1. Here, \(J\) represents the coupling constant and \(D\) denotes the single-spin anisotropy parameter. A key advantage of this model is its ability to capture both first-order and second-order phase transitions simultaneously. The model transitions from a second-order to a first-order phase transition as \(D\) increases. The widely recognized transition point for this change is \(D/J=1.965\) and \(k_B T/J = 0.609\) \cite{silva2006wang}, which is commonly known as the tricritical point. For simplicity, we set the Boltzmann constant \(k_B = 1\), and similarly set \(J = +1\) for the remainder of our study.

\subsection{Wang-Landau Sampling}

    The Wang-Landau (WL) algorithm \cite{landau2004new} is a highly efficient and broadly applicable method for directly estimating the density of states, $g(E)$, for a system. Utilizing $g(E)$ obtained through this method enables the direct investigation of higher-order phase transitions while avoiding additional processing, thereby minimizing errors and enhancing result accuracy. The density of states $g(E)$ is estimated through a random walk in energy space, with the histogram flattening as a key criterion, using the Wang-Landau (WL) sampling method. As the exact values of the density of states are unknown at the beginning of the simulation, all entries are initially set to $g(E) = 1$ and subsequently refined according to a specified probability distribution.
    \begin{equation}
    p(E_i \rightarrow E_j) = \min\left(\frac{g(E_i)}{g(E_j)}, 1\right),
	\end{equation}
    where $E_i$ and $E_j$ are the energies before and after a spin flip, the density of states is modified by
    \begin{equation}
    g(E) \rightarrow g(E)f,
	\end{equation}
    where $E$ represents the energy level of the accepted state and $f$ is the modification factor, initially set to $f = f_0 = e = 2.71828$. Simultaneously, the energy histogram $H(E)$ is incremented by 1. The random walks continue, flipping spin values in the lattice configuration to generate new energy values for each iteration, until the energy histogram becomes ``flat''. A flat histogram is defined as the minimum value of any energy level's histogram being at least 80\% of the average value. The modification factor is then reduced to $f_{i+1} = \sqrt{f_i}$, and the histogram is reset to zero for the next iteration process. The simulation concludes when $f$ falls below the threshold of $f_{\text{final}} = 1 + 10^{-8}$.
    
    During the sequential simulation process, when the value of $D$ exceeds 1.9 (approaching a first-order phase transition), we encountered a problem in which a significant portion of the simulations only explored the negative energy levels, while only a few simulations captured the entire range of energy levels. This result clearly did not align with physical expectations, leading to significant inaccuracies in the simulation outcomes. 
    As the system size increases and more information is revealed, metastable states emerge, making it difficult for serial simulations to escape the process, ultimately causing the simulation to display only negative energy levels. To address this issue, we utilized two cores for parallel computing, assigning each core a portion of the energy levels, with a 0.75 overlap between the portions managed by the two cores. Additionally, we modified the initialization method of the two-dimensional lattice. At the start of the simulation, instead of initializing the entire array with a value of 1, we initialized the energy levels of the lattice within the range of one-third to two-thirds of each core's responsibility. Upon the completion of the simulation, we merged the data obtained from both cores to achieve a complete density of states.
    \begin{figure}
        \centering 
		\epsfig{figure=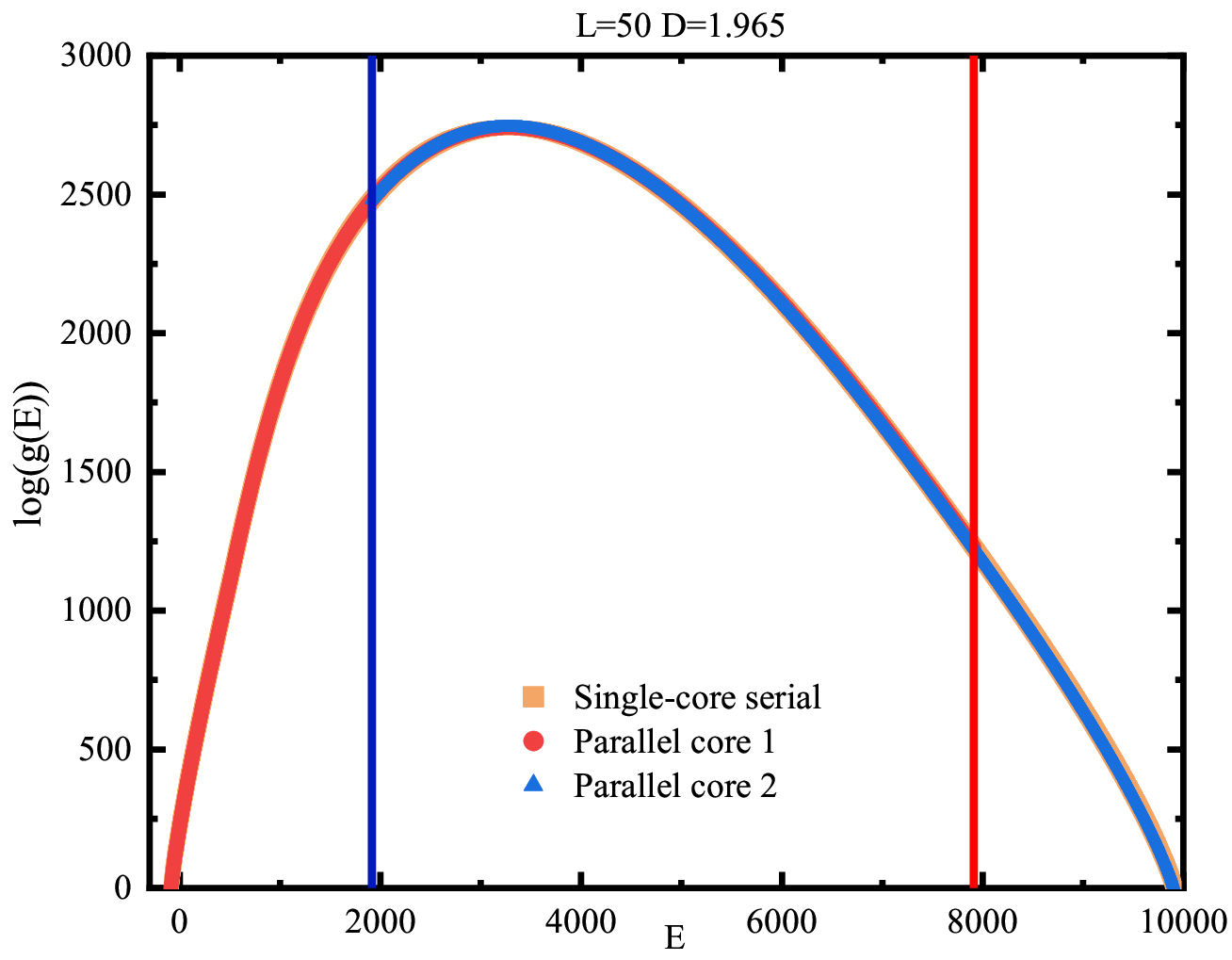,width=0.5\linewidth} \caption{(Algorithm Comparison Graph).
        The density of states obtained from two different algorithms was compared. To improve clarity, the data from the serial calculation is displayed with slightly larger symbols and highlighted at the base using light yellow markers. The data computed in parallel on two cores is represented by red and blue dots. Additionally, vertical red and blue lines were added to indicate the energy level boundaries calculated by each core. These adjustments enhance the visualization, making it easier to observe the overlapping energy levels and their distribution, thereby improving the readability and comparative effect of the results.
		}
         \label{Algorithm}
   \end{figure}
    In Fig. \ref{Algorithm}, we present the comparison between the two algorithms. The data from parallel computation on two cores and serial computation on a single core perfectly overlap, confirming the correctness of our algorithm.
    
\subsection{Microcanonical infection-point analysis method}
	Qi and Bachmann integrated microcanonical analysis with the principle of minimum sensitivity to identify and classify first-order and higher-order transitions in complex systems \cite{qi2018classification}. Their analysis revealed that the phase transition signals from the two-dimensional ferromagnetic Ising model indicated not only straightforward second-order transitions but also more complex behaviors. Moreover, both independent and dependent third-order transitions were observed in the system. 

    In a physical system, macroscopic behavior is governed by entropy and energy, while microcanonical entropy, which encapsulates all information about the system's phase behavior, is defined as follows:
    \begin{equation}
		S(E) = k_B \ln g(E).
	\end{equation}
    Here, $k_B$ denotes the Boltzmann constant, while $S(E)$ represents entropy; its derivatives maintain consistent concavity within the energy region associated with a single phase. However, phase transitions disrupt this consistent concavity, resulting in an evident inflection point in the graph. According to the principle of minimal sensitivity, only the least sensitive inflection points hold physical significance.
    If $S(E)$ exhibits a least sensitive inflection point, then its first derivative is given by:
    \begin{equation}
    \beta(E) = T^{-1}(E) = \frac{dS(E)}{dE},
    \end{equation}
    where $\beta(E)$, representing the microcanonical inverse temperature, is expected to exhibit a positive minimum. This indicates a first-order phase transition. If $\beta(E)$ also displays a least sensitive inflection point, we can further analyze it by obtaining its first derivative:
    \begin{equation}
    \gamma(E) = \frac{d\beta(E)}{dE} = \frac{d^2S(E)}{dE^2},
    \end{equation}
    where $\gamma(E)$, describes the rate of change of the microcanonical inverse temperature, and provides insights into first-order and second-order phase transitions. In addition,
    \begin{equation}
    \delta(E) = \frac{d\gamma(E)}{dE} = \frac{d^3S(E)}{dE^3}.
    \end{equation}
    Here, $\delta(E)$, representing the third derivative of $S(E)$ with respect to $E$, can provide insights into third-order phase transitions. This allows for the distinction between both independent and dependent third-order transitions. Building upon this foundation, we also calculated the fourth derivative of $S(E)$, denoted as $\epsilon(E)$ in this paper. On one hand, this enables a more in-depth analysis of the characteristics of the third derivative; on the other hand, we explore the potential for detecting higher-order signals. By studying the fourth derivative, we aim to uncover more subtle changes in the system's energy states, thereby enhancing our understanding of the phase transition mechanism. Using this method, further classification can be obtained in Table \ref{Signal}.
    \begin{table}[h!]
    \centering
    \caption{Signal of the order of the transitions}
    \begin{tabular}{ccc}
    \hline\hline
    {Categories} & {Even order transitions} & {Odd order transitions} \\
    \hline
    \multirow{2}{*}{Independent} & $\frac{d^{2k}S(E)}{dE^{2k}} < 0$ & $\frac{d^{2k-1}S(E)}{dE^{2k-1}} > 0$ \\
    & Negative maximum & Positive minimum \\
    \hline
    \multirow{2}{*}{Dependent} & $\frac{d^{2k}S(E)}{dE^{2k}} > 0$ & $\frac{d^{2k-1}S(E)}{dE^{2k-1}} < 0$ \\
    & Positive minimum & Negative maximum \\
    \hline\hline
    \end{tabular}
    \label{Signal}
    \end{table}
    
    Derivatives obtained directly from discrete microcanonical entropy are influenced by numerical noise present in the data. To minimize this noise, we processed the data using a denoising procedure. Initially, we computed the density of states (DOS) five times and averaged the results to obtain a smoother curve. Subsequently, we applied the Bézier algorithm to generate a smooth function \cite{bachmann2014thermodynamics}. This was also mentioned in our previous work. Repeating the above steps, we obtained four sets of data and calculated the error bars.
    
    \subsection{Geometrical analysis methods}
    To attain a more comprehensive understanding of this model, we conducted a further analysis from a geometric perspective. The spin configuration of the system was generated using Metropolis sampling \cite{metropolis1953equation}. The geometric order parameter, defined to characterize the system's phase transition, was subsequently extracted from the configuration data. The first order parameter we used is the isolated spin. Our definition of isolated spins differs slightly from that in the work of Sitarachu and Bachmann \cite{sitarachu2022evidence}. In this study, an isolated spin is defined by two conditions: first, the spin's state must differ from all its nearest neighbors; second, these neighboring spins must belong to the same cluster. Only when both conditions are met can the spin be classified as an isolated spin. The role of isolated spins is to disrupt both ordered and disordered states; thus, they may also be referred to as ``disruptors''. In the vicinity of the third-order independent transition, the number of isolated spins tends to increase. In this paper, we denote $I_{SO}$ to represent isolated spins. 
    The second quantity is the average perimeter of clusters. Initially, we attempted to study the average cluster size, but we were unable to extract information related to the third-order dependent transition.
    We hypothesized that analyzing information from different dimensions within the same system would provide distinct insights into the phase transition. However, given that our model is constrained to a two-dimensional plane, we were unable to increase its dimensionality. As a result, we chose to apply dimensional reduction, focusing on the average perimeter of clusters. Using this approach, we successfully extracted signals associated with the third-order dependent transition.
    
    We define $\Pi$ as the average perimeter of clusters composed of more than one spin within a specific spin configuration $X$. Specifically, for a spin configuration $X$, we first identify all clusters composed solely of more than one spin and then calculate the average perimeter of these clusters. 
    \begin{equation}
     \Pi = \frac{1}{n} \sum_{l} C_{l}.
    \end{equation}
    In this context, $n$ represents the total number of clusters in $X$ that contain more than one spin, $l$ denotes the clusters with more than one spin, and $C_l$ refers to the number of spins in cluster $l$. Thus, the statistical average obtained is
    \begin{equation}
     \langle \Pi \rangle = \frac{1}{Z} \sum_X \Pi(X) e^{-E(X)/k_B T},
    \end{equation}
    where $T$ is the canonical temperature and $Z = \sum_X \exp\left[-E(X)/k_B T\right]$ is the canonical partition function. However, our primary interest lies in the rate of change of this quantity. Consequently, we performed a first-order derivative calculation to obtain its rate of change. We denote this quantity as $D_{\langle \Pi \rangle}$, which represents its rate of change, expressed as:
    \begin{equation}
     D_{\langle \Pi \rangle} =  \frac{d\langle \Pi \rangle}{dT}.
    \end{equation}
    When a peak in $D_{\langle \Pi \rangle}$ is observed, we define its position as the critical point. 
    The region beyond the critical point is termed the disordered region or the high-temperature region. If a local minimum is observed in this region, we identify it as the location of the third-order dependent transition. This definition is consistent with the one used by Sitarachu and Bachmann in their article \cite{sitarachu2022evidence} for assessing the average cluster size. Through further analysis and subsequent results, we have verified the accuracy and applicability of this definition, ensuring its reliability across different system parameters and conditions.

    \section{Results}
    \subsection{Traditional transitions}
	
\begin{figure}
		\epsfig{figure=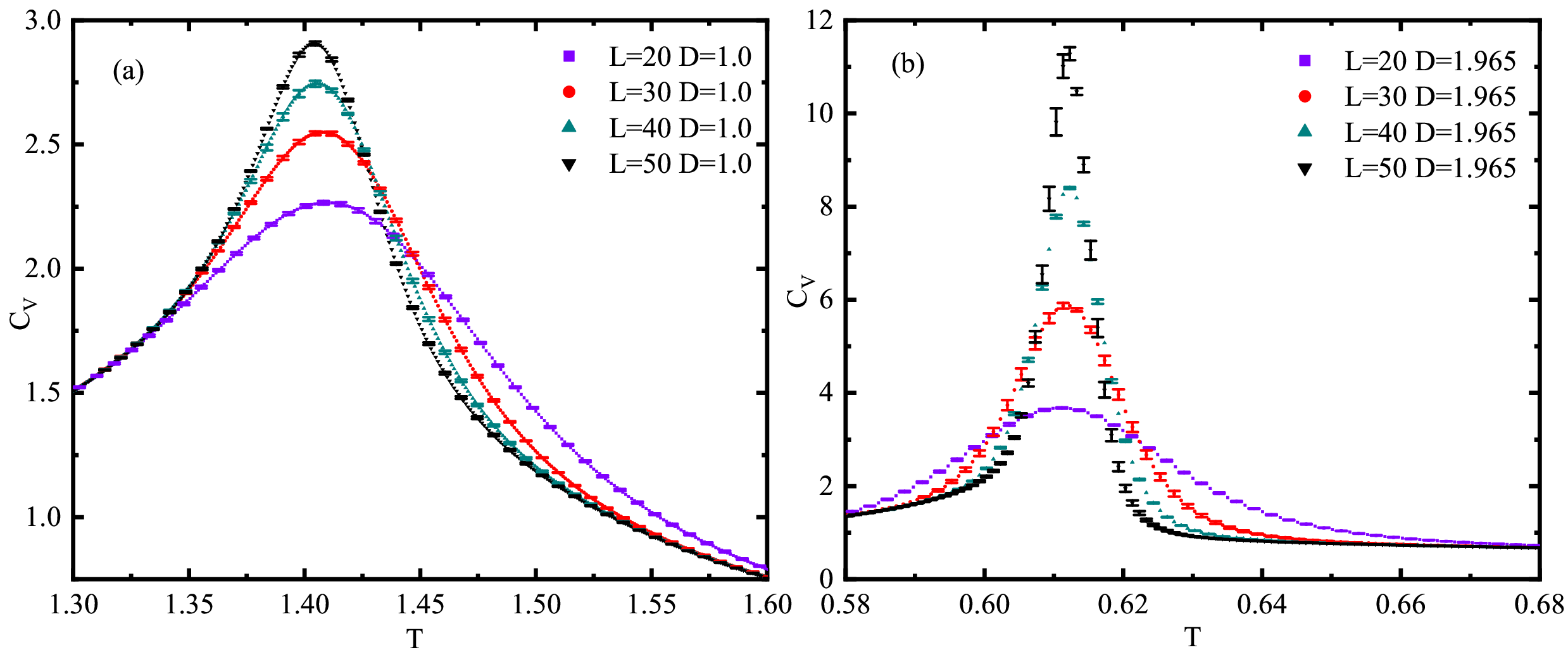,width=1.0\linewidth} \caption{(Specific heat graph obtained under the canonical ensemble).
       The specific heat calculated at various \(D\) values is presented here, with the peak of the specific heat serving as an indicator of the phase transition point. Figure (a) shows the specific heat analysis for $D = 1.0$, while Figure (b) illustrates the specific heat analysis for $D = 1.965$. Different colors and markers are used to distinguish various system sizes.
		}
         \label{Specific_heat}
   \end{figure}

\begin{figure}
		\epsfig{figure=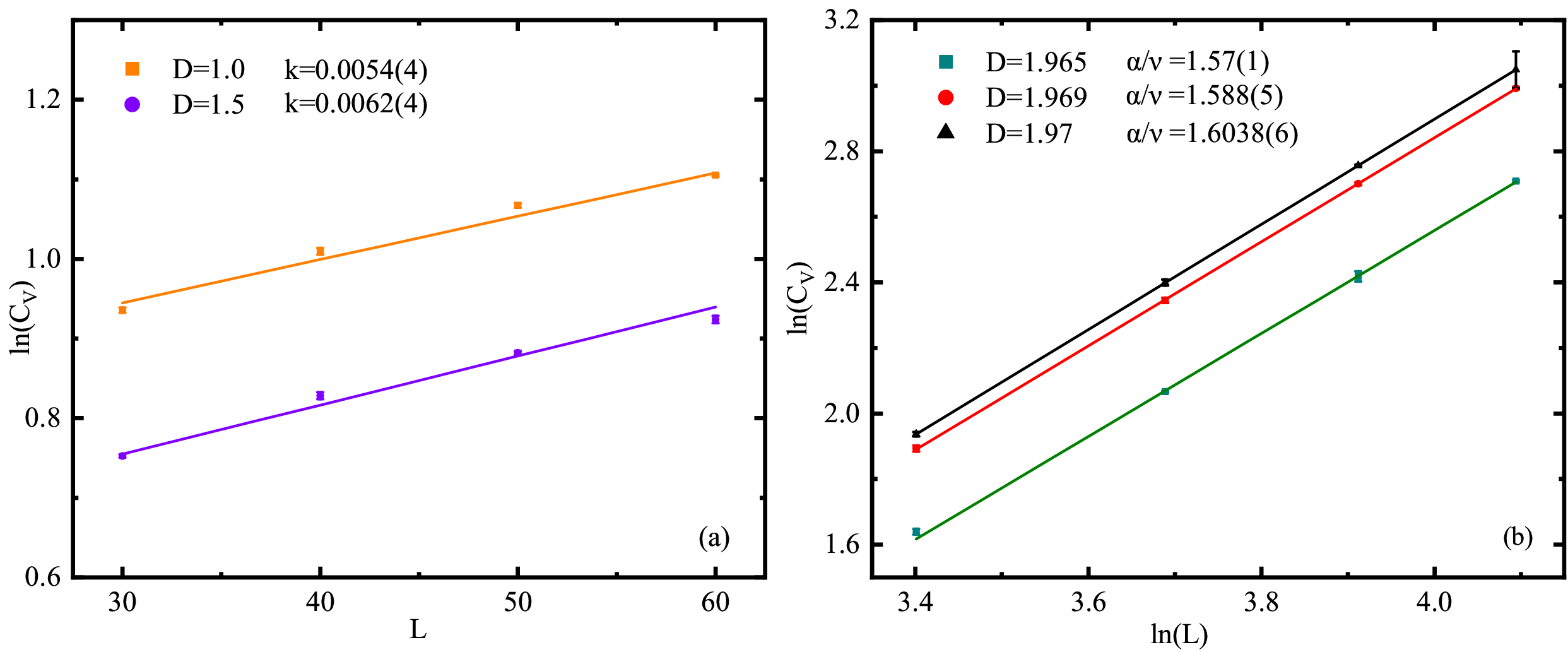,width=1.0\linewidth} \caption{(Finite size analysis of the Blume-Capel Model).
        Figure (a) displays a semi-logarithmic analysis conducted when the primary phase transition of the system is of the second order, revealing that its critical behavior follows logarithmic divergence, where $k$ represents the extracted slope. Figure (b) illustrates a log-log analysis performed when the primary phase transition is of the first order, incorporating the extracted critical exponent into the legend, where the tricritical point is characterized by a critical exponent of $\alpha/\nu = 1.57$.
		}
         \label{FFS}
   \end{figure}

	In this study, we simulated systems of sizes $L = 20, 30, 40, 50,$ and $60$ under various values of $D$.  We obtained the density of states data through Wang-Landau (WL) sampling and identified the phase transition points using canonical methods and microcanonical inflection point analysis. The specific heat of the system was calculated using the following formula:
    \begin{equation}
		C_V = \frac{\langle E^2 \rangle - \langle E \rangle^2}{N k_B T^2},
		\label{C_V}
    \end{equation}
    where $\langle E^n \rangle = \sum_E E^n g(E) e^{-E/k_B T}$ and $N = L \times L$. We divide the specific heat by \(N\) to enable the plotting of specific heat graphs for different sizes on the same chart, thereby facilitating our observation. In Fig. \ref{Specific_heat}, we present a plot determining the phase transition points using specific heat for \(D = 1.0\) and \(D = 1.965\) in the canonical ensemble. We clearly observe that the specific heat exhibits a peak, with the temperature corresponding to this peak indicating the location of the phase transition point. To verify the accuracy of our data, we conducted a finite-size analysis. However, during the simulation, we approximated the energy level spacing to be 1, which resulted in the loss of some detailed information. 
    Consequently, the specific heat was adjusted using the following formula \cite{fernandes2015critical} when the primary phase transition of the system is of the first order: 
    \begin{equation}
		C_v = aL^{\alpha/\nu} \left(1 + bL^{-\omega}\right).
		\label{C_V_xiuzheng}
    \end{equation}
    Here, $\omega$ denotes the primary correction exponent, while $a$ and $b$ are non-universal constants. This correction enhanced the accuracy of the specific heat in the first-order phase transition, thereby increasing the reliability of the results. When the primary phase transition of the system is of the second order, the classical 2D Blume-Capel model is categorized within the same universality class as the 2D Ising model. Consequently, we conducted a single-logarithmic analysis, as shown in Figure \ref{FFS}(a), which indicates that the system exhibits logarithmic divergence. However, during first-order phase transitions, the critical behavior markedly deviates, and the system ceases to adhere to the universality class of the 2D Ising model. According to Ref. \cite{moueddene2024critical}, the expected critical exponent at the tricritical point is $\alpha/\nu = 1.60$. The figure in Ref. \cite{kwak2015first} shows that the critical exponent near the tricritical point is approximately $\alpha/\nu \approx 1.608$. These reference values provide important theoretical support for our analysis. In this study, when $D = 1.965$, the critical exponent obtained from Fig. \ref{FFS}(b) was $\alpha/\nu = 1.57(1)$, which is in close agreement with the reference values, further validating the accuracy of our data and the reliability of our analytical methods.
    
\begin{figure}
		\epsfig{figure=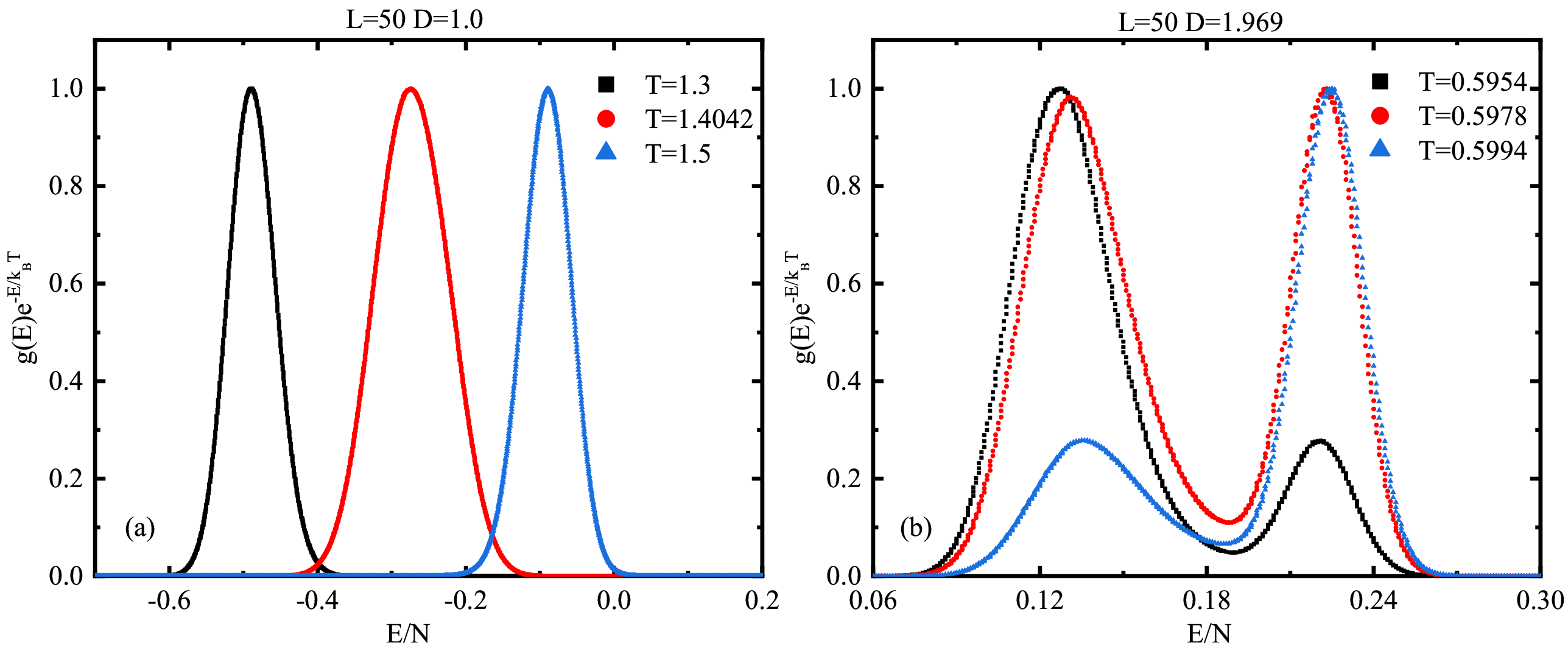,width=1.0\linewidth} \caption{(For the Blume-Capel model with $L = 50$, the canonical distribution $P(E, T) \sim g(E) e^{-E/k_B T}$ is examined at the transition temperature $T$).
        (a) For $D = 1.0$, the distribution of $P$ displays a single-peak form, suggesting that the primary phase transition is of the second order. (b) For $D = 1.965$, the distribution of $P$ shows a double-peak form, implying that the primary phase transition is of the first order. The red points represent data corresponding to the phase transition temperature, the black points represent data slightly below it, and the blue points represent data slightly above it.
		}
         \label{P}
   \end{figure}

    To more thoroughly validate our assessment of the type of phase transition, we present the canonical distribution $P(E, T) \sim g(E) e^{-E/k_B T}$ near the phase transition temperature, as illustrated in Fig. \ref{P}. In Fig. \ref{P}(a), the canonical distribution exhibits a single-peaked shape, clearly indicating that the phase transition at this point is of second order. As the temperature increases, the peak position shifts to higher energy levels, and the energy level corresponding to the peak is precisely the energy level at which the system undergoes the phase transition. However, in Fig. \ref{P}(b), the canonical distribution displays a double-peaked shape, clearly indicating that the phase transition at this point is of first order. In contrast to the second-order phase transition, the phase transition energy level of a first-order transition is located at the valley between the two peaks. The double-peak structure reflects phase coexistence in a first-order phase transition, where the energy level corresponding to the valley represents the phase transition energy level. This observation aligns with the transition energy level identified through subsequent microcanonical inflection-point analysis.
   
     \begin{table}[ht!]
     \centering
     \caption{Microcanonical and canonical analysis results for different D values}
     \begin{tabular}{ccccccc} 
     \hline\hline
     $D$ & & $L=20$ & $L=30$ & $L=40$ & $L=50$ & $L=60$ \\
     \hline
     \multirow{3}{*}{$1.0$}& $1/\beta_c$ & 1.4097(0.0004) & 1.4079(0.0004) & 1.4055(0.0005) & 1.4042(0.0002) & 1.4036(0.0001) \\
     &$1/\beta_m$ & 1.4281 & 1.4178 & 1.4113 & 1.4085 & 1.4067 \\
     &$E/N$& -0.2625 & -0.2589 & -0.2631 & -0.2644 & -0.2658 \\
     \hline
     \multirow{3}{*}{$1.5$} & $1/\beta_c$ & 1.1569(0.0009) & 1.1571(0.0002) & 1.1564(0.0003) & 1.1555(0.0003) & 1.1551(0.0002)\\
     &$1/\beta_m$ & 1.1743 & 1.1661 & 1.1614 & 1.1592 & 1.15801 \\
     &$E/N$& 0.0950 & 0.09889 & 0.0975 & 0.0968 & 0.09667 \\
     \hline
     \multirow{3}{*}{$1.965$} & $1/\beta_c$ & 0.6110(0.0001) & 0.6118(0.0003) & 0.6122(0.0001) & 0.6120(0.0001) & 0.6120(0.0001) \\
     & $1/\beta_m$ & 0.6171 & 0.6152 & 0.6140 & 0.6129 & 0.6122 \\
     & $E/N$ & 0.1725 & 0.1833 & 0.1944 & 0.1976 & 0.2022 \\
     \hline
     \multirow{3}{*}{$1.969$}& $1/\beta_c$ & 0.5975(0.0005) & 0.5983(0.0003) & 0.5980(0.0003) & 0.5978(0.0001) & 0.5978(0.0001) \\
     & $1/\beta_m$ & 0.6171 & 0.6051 & 0.5995 & 0.5980 & 0.5967 \\
     & $E/N$ & 0.1725 & 0.1533 & 0.1800 & 0.1832 & 0.1875 \\
     \hline
     \multirow{3}{*}{$1.97$} & $1/\beta_c$ & 0.5940(0.0034) & 0.5947(0.0002) & 0.5942(0.0004) & 0.5938(0.0002) & 0.5937(0.0001) \\
     & $1/\beta_m$ & 0.5995 & 0.5979 & 0.5958 & 0.5938 & 0.5926 \\
     & $E/N$ & 0.1575 & 0.1700 & 0.1756 & 0.1976 & 0.1825 \\
     \hline\hline
     \end{tabular}
     \label{TC}
     \end{table}

    Additionally, we conducted a microcanonical inflection point analysis on the data. In Figs \ref{1_0}(b), \ref{1_965}(b), and \ref{1_97}(b), we similarly determined the critical and first-order phase transition points. In Table \ref{TC}, we compile the phase transition point information determined through microcanonical inflection point analysis and the heat capacity peak method. $1/\beta_c$ represents the phase transition temperature calculated in the canonical ensemble, while $1/\beta_m$ represents the phase transition temperature obtained using the microcanonical method. In the last column of the table, we have also added the energy level position $E/N$ where the phase transition occurs, and the numbers in parentheses indicate the error range of our calculations. Through this information, we can more clearly observe the accuracy and differences between the phase transition temperatures and their corresponding energy levels. A detailed comparison reveals that the phase transition temperatures obtained from the two methods are highly consistent, demonstrating the reliability and accuracy of our calculated results. Furthermore, when we compare our results with the phase transition points mentioned in the literature Refs. \cite{silva2006wang, kwak2015first} and \cite{beale1986finite}, we find that the numerical values are very close.
	
    \subsection{Higher order transitions}
\begin{figure}
        \centering 
		\epsfig{figure=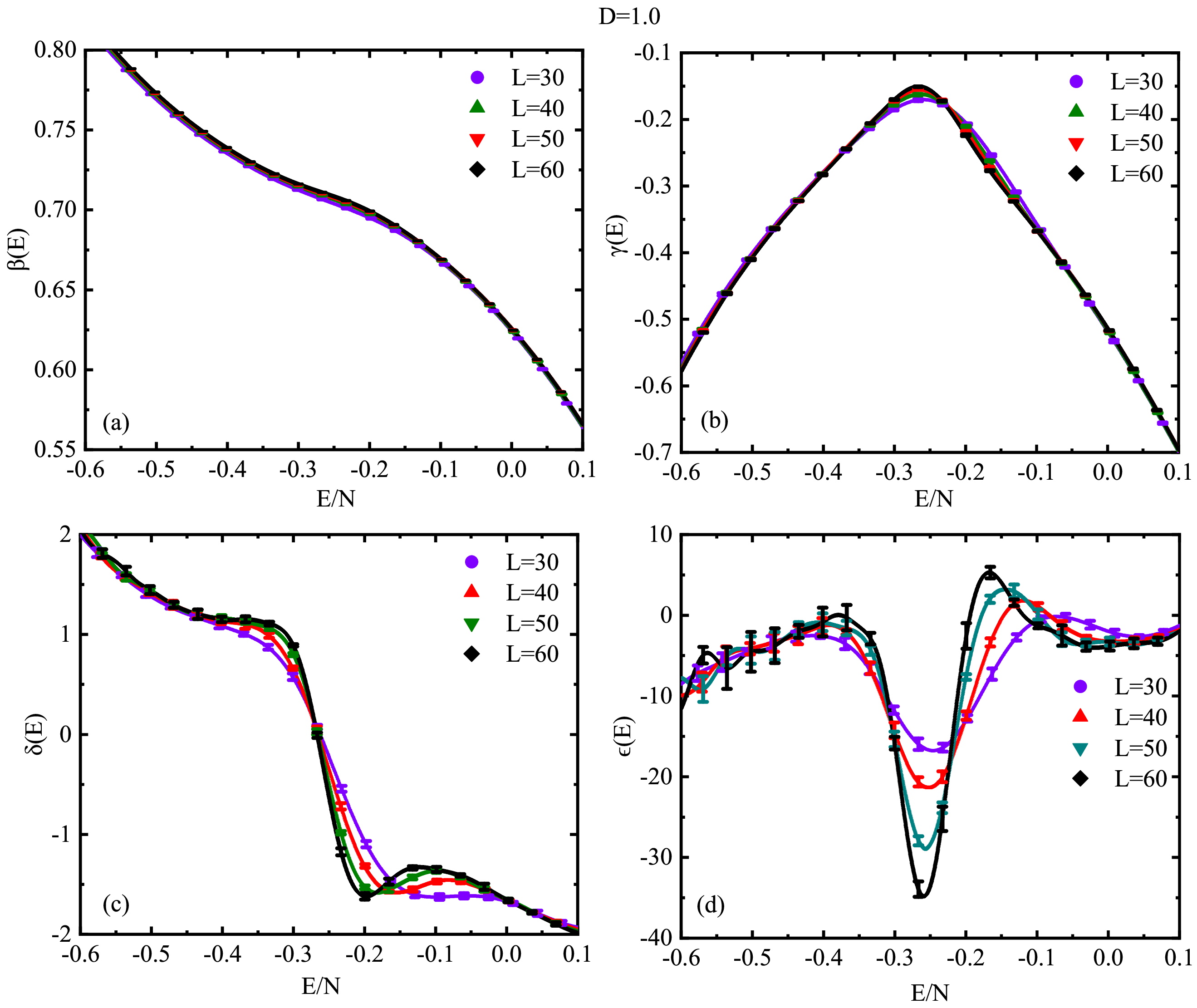,width=0.60\linewidth} \caption{(Microcanonical inflection point analysis of the derivative data for D=1.0).
	   The detailed results for the Blume-Capel model within the microcanonical ensemble at $D = 1.0$ are presented. Figure (a) shows the variation of the first derivative of entropy, $\beta(E)$, while Figure (b) depicts the trend of the second derivative, $\gamma(E)$. Figure (c) highlights the evolution of the third derivative, $\delta(E)$, with a positive minimum and a negative maximum identified as indicators of third-order independent and dependent transitions, and Figure (d) reveals the changes in the fourth derivative, $\epsilon(E)$. 
		}
            \label{1_0}
        \end{figure}

\begin{figure}
        \centering
		\epsfig{figure=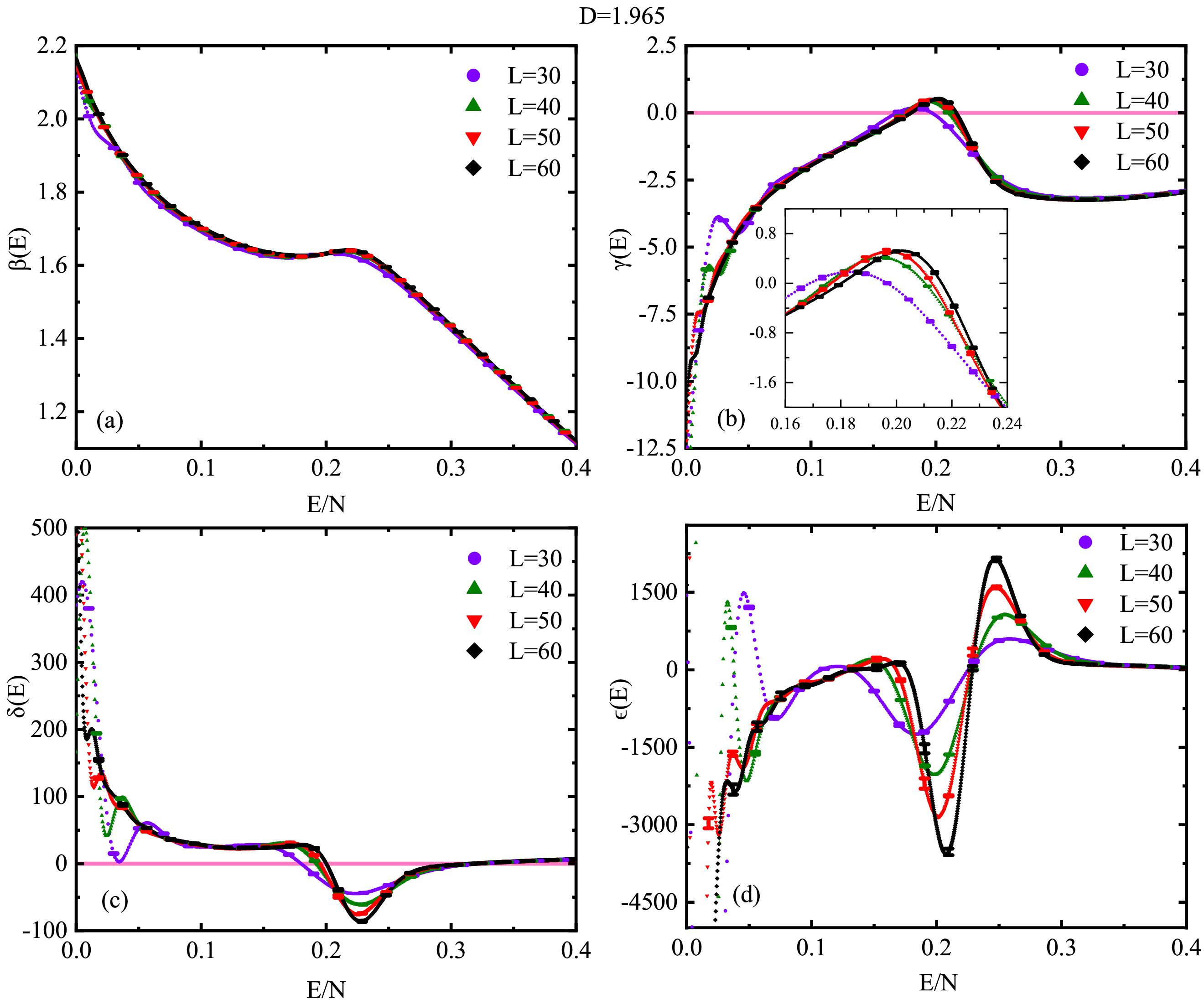,width=0.60\linewidth} \caption{(Microcanonical inflection point analysis of the derivative data for D=1.965).
        The detailed results for the Blume-Capel model within the microcanonical ensemble at $D = 1.965$ are presented. Figure (a) shows the variation of the first derivative of entropy, $\beta(E)$, while Figure (b) depicts the trend of the second derivative, $\gamma(E)$. Figure (c) highlights the evolution of the third derivative, $\delta(E)$, with a positive minimum at higher energy levels identified as an indicator of a third-order independent transition, and Figure (d) reveals the changes in the fourth derivative, $\epsilon(E)$. 
		}
            \label{1_965}
        \end{figure}

\begin{figure}
       \centering
		\epsfig{figure=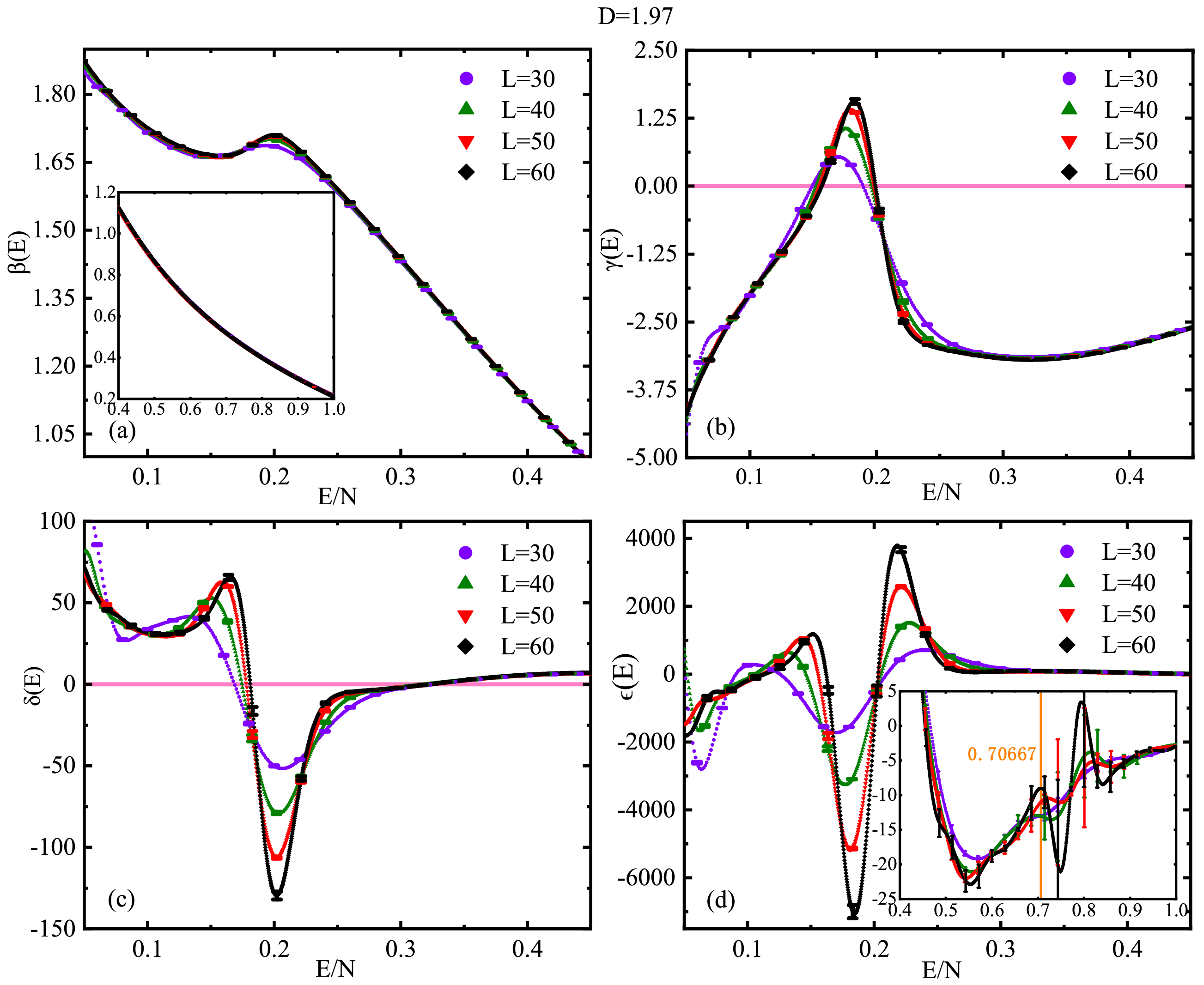,width=0.60\linewidth} \caption{(Microcanonical inflection point analysis of the derivative data for D=1.97).
			The detailed results for the Blume-Capel model within the microcanonical ensemble at $D = 1.97$ are presented. Figure (a) shows the variation of the first derivative of entropy, $\beta(E)$, with an inset depicting its extended part. Figure (b) depicts the trend of the second derivative, $\gamma(E)$. Figure (c) illustrates the evolution of the third derivative, \(\delta(E)\), with the positive minimum identified as an indicator of a third-order independent transition, and Figure (d) showcases the changes in the fourth derivative, \(\epsilon(E)\), with an inset showing its extended part and an orange vertical line indicating the identified position of the fourth-order independent transition.
		}
            \label{1_97}
        \end{figure}

	We use the $\delta(E)$ curve to identify third-order phase transition points. In Figs. \ref{1_0}, \ref{1_965}, and \ref{1_97}, we present the derivative graphs of the density of states at various orders for different system sizes and $D$ values. To ensure data accuracy and improve visualization, error bars were added to every fixed number of data points in the figures. From these graphs, we can observe that the type of phase transition changes with varying $D$ values, and we can also identify the occurrence of higher-order phase transitions. Identifying higher-order transition types helps further analyze and optimize data utilization, thereby enhancing our overall understanding of the system.
 
    In Fig. \ref{1_0}, we present the analysis for \(D = 1.0\). Figures \ref{1_0}(a)-(d) respectively display the first, second, third, and fourth derivatives of the density of states. In Fig. \ref{1_0}(a), a least-sensitive inflection point is observed, indicating the phase transition point under these conditions. In Fig. \ref{1_0}(b), a significant negative peak is observed in the region, indicating that the phase transition is of second order. In Fig. \ref{1_0}(c), we can observe a positive minimum and a negative maximum over the entire interval, representing signals for independent and dependent third-order transitions, respectively. Based on the analysis of Figure \ref{1_0}, we conclude that when $D = 1.0$, the primary phase transition in the system is of second order, and the signals for third-order transitions, along with their corresponding temperature and energy information, are also identified.
    
    In Fig. \ref{1_965}, we conduct an in-depth analysis of the region near the tricritical point. A reference line at $y = 0$ is added to Figs. \ref{1_965} (b) and (c) to clarify the trends. In Fig. \ref{1_965}(b), we observe a positive peak, indicating that the primary phase transition of the system is first-order. The observed peak value of \(\gamma\) is relatively small, suggesting that the system's primary phase transition recently shifted from second-order to first-order. In Fig. \ref{1_965} (c), we observe two distinct positive minima on the left side of the transition point. However, as the system size increases, the minimum near the ground-state energy flattens and shows signs of disappearing. This suggests that the appearance of this point is likely due to noise in the simulation and boundary effects. The fluctuations observed in Fig. \ref{1_965} (d) further support this hypothesis, providing strong evidences for our interpretation. Therefore, the minima observed at higher energy levels are identified as indicators of third-order independent transitions.
        
    In Fig. \ref{1_97}, there is a certain distance from the tricritical point. As shown in Fig. \ref{1_97}(b), the height of the positive peak at this point is significantly higher than the peak in Fig. \ref{1_965}, indicating that the phase transition characteristics of the system are more pronounced. Similar to Figs. \ref{1_965}(c) and \ref{1_97}(c), we observe only a positive minimum and do not detect any signal of a negative maximum. 
    This suggests that when the primary phase transition of the system is of the first order, there is no signal for a third-order dependent transition. However, our exploration of the high-temperature region has revealed a signal for a fourth-order independent transition, indicated by an orange vertical line in the inset of Figure \ref{1_97}(d), complemented by a corresponding inset in Figure \ref{1_97}(a) to facilitate temperature determination. 
    By comparing Figs. \ref{1_0}(b), \ref{1_965}(b), and \ref{1_97}(b), we observe that as the value of $D$ and system size $L$ increase, the peak of $\gamma(E)$ increases progressively but eventually converges to a fixed value. In fact, $\gamma(E)$ is related to a thermodynamic property associated with the specific heat phase transition. It is observed that for $D = 1.0$, the peak of $\gamma(E)$ changes very little with system variation, whereas for $D = 1.97$, the peak changes significantly but eventually approaches a fixed value. This occurs because a first-order phase transition is accompanied by latent heat, and as the system size increases, the latent heat gradually approaches the thermodynamic limit, eventually converges to a stable value. We summarize the analysis results for all parameters in Table \ref{third order transitions}. In the table, $1/\beta_d$ represents the temperature of the third-order dependent transition, while $1/\beta_i$ denotes the temperature of the third-order independent transition. ``NF'' indicates cases where no signal was found. The corresponding energy levels are also provided in the table to clearly illustrate the correlation between each signal and the system's energy states.
     \begin{table}[ht!]
     \centering
     \setlength{\tabcolsep}{16pt} 
     \caption{Third-order transition positions in the Blume-Capel model from microcanonical analysis}
     \begin{tabular}{ccccccc} 
     \hline\hline
     $D$ & & $L=20$ & $L=30$ & $L=40$ & $L=50$ & $L=60$ \\
     \hline 
     \multirow{4}{*}{$1.0$}& $1/\beta_d$ & NF & 1.7719 & 1.4780 & 1.4962 & 1.4748 \\
     &$E/N$& - & -0.0733 & -0.1194 & -0.0972 & -0.1242 \\
     &$1/\beta_i$ & 1.3366 & 1.3512 & 1.3605 & 1.3512 & 1.3641\\
     &$E/N$& -0.4650 & -0.4167 & -0.3919 & -0.4068 & -0.3792 \\
     \hline
     \multirow{4}{*}{$1.5$} & $1/\beta_d$ & NF & NF & 1.2580 & 1.2628 & 1.2336 \\
     &$E/N$& - & - & 0.2375 & 0.2324 & 0.2164 \\
     &$1/\beta_i$ & 1.0811 & 1.0956 & 1.1046 & 1.1179 & 1.1139 \\
     &$E/N$& -0.0675 & -0.0356 & -0.0175 & 0.0068 & 0.0011 \\
     \hline
     \multirow{4}{*}{$1.965$} & $1/\beta_d$ & NF & NF & NF & NF & NF \\
     & $E/N$ & - & - & - & - & - \\
     & $1/\beta_i$ & NF & 0.5940 & 0.6116 & 0.6041 & 0.6035 \\
     & $E/N$ & - & 0.1044 & 0.1475 & 0.1300 & 0.1308 \\
     \hline
     \multirow{4}{*}{$1.969$}& $1/\beta_d$ & NF & NF & NF & NF & NF \\
     & $E/N$ & - & - & - & - & - \\
     & $1/\beta_i$ & NF & 0.5935 & 0.5908 & 0.5936 & 0.5913 \\
     & $E/N$ & - & 0.1156 & 0.1144 & 0.1208 & 0.1181 \\
     \hline
     \multirow{4}{*}{$1.97$} & $1/\beta_d$ & NF & NF & NF & NF & NF \\
     & $E/N$ & - & - & - & - & - \\
     & $1/\beta_i$ & 0.5640 & 0.5569 & 0.5855 & 0.5885 & 0.5890 \\
     & $E/N$ & 0.0775 & 0.0833 & 0.1075 & 0.1144 & 0.1164 \\
     \hline\hline
     \end{tabular}
     \label{third order transitions}
     \end{table}
        
    To further investigate the underlying mechanisms of higher-order transitions, we conducted an additional analysis from a geometric perspective. Since geometric properties may not be clearly exhibited in small systems, we performed simulations with system sizes $L = 30, 40, 50, 60$ and values of $D = 1.0, 1.5, 1.965, 1.97$. These parameters were selected to ensure that the relevant geometric features could be adequately captured and analyzed.
    \begin{figure}
    \centering
    \epsfig{figure=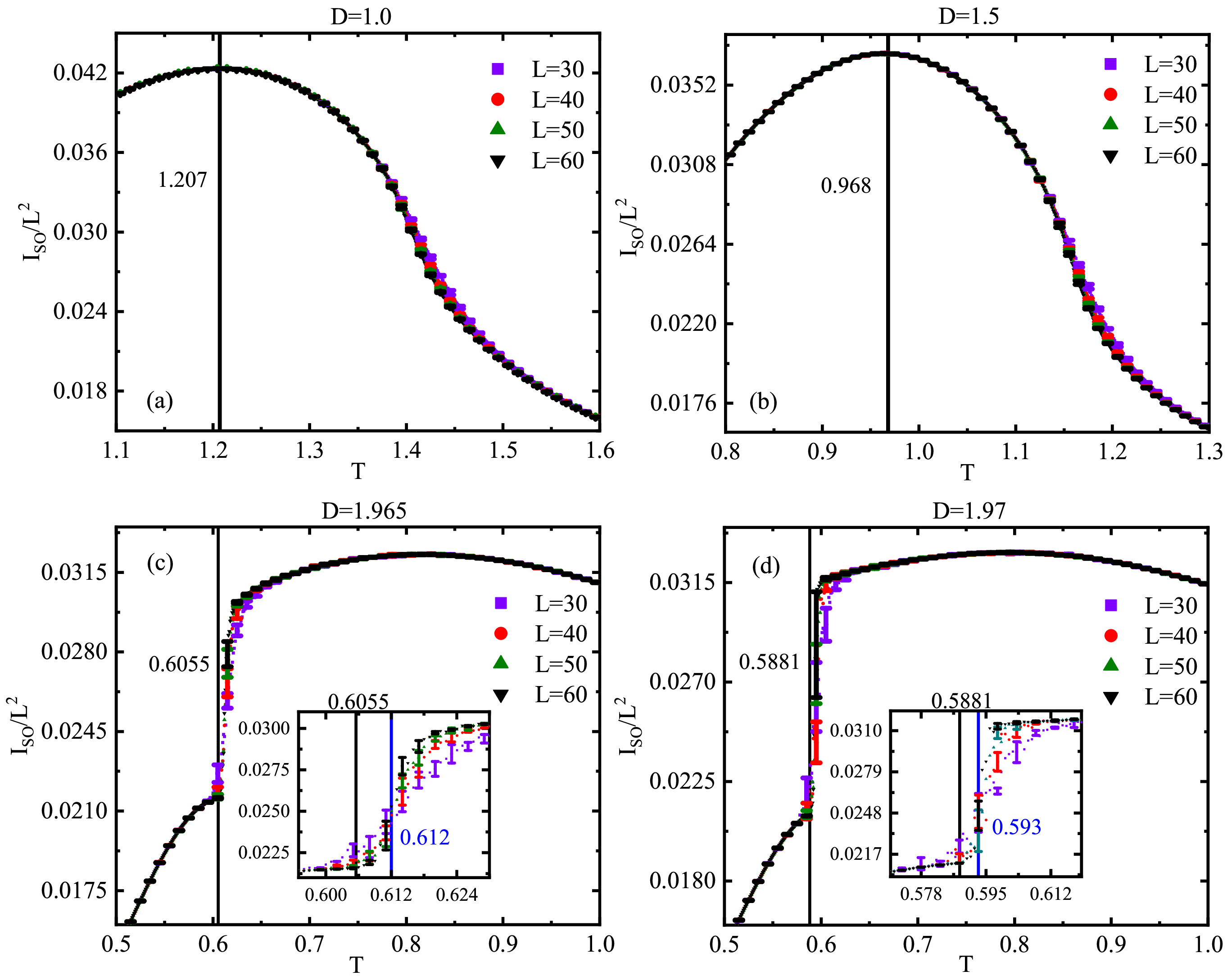,width=0.60\linewidth} \caption{(Images of isolated spins under different parameters).
    Here, the variations in isolated spins across different parameters are presented. Black vertical lines in the figures mark the positions of the third-order independent transitions, with the corresponding temperatures labeled in the same color next to them. The insets in Figures (c) and (d) represent magnified views of the discontinuous jumps in isolated spins, where blue vertical lines indicate the phase transition points and the corresponding temperatures.
    }
        \label{ISO}
    \end{figure}
    \begin{figure}
    \epsfig{figure=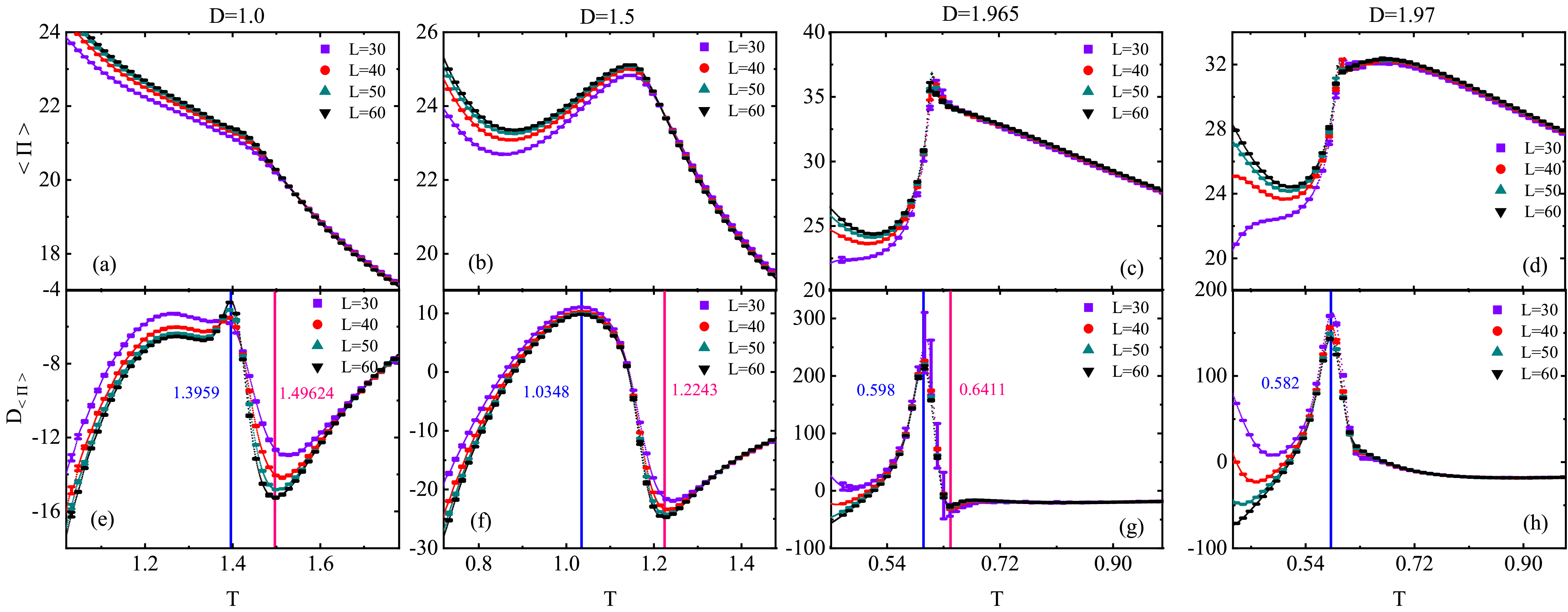,width=1.0\linewidth} \caption{(The data of the average perimeter under different parameters and the variations of its first derivative).
    Here, the variations of the average perimeter and its first derivative across different parameters are presented. Blue vertical lines have been added at the positions where the critical points are identified, and pink vertical lines mark the positions of the third-order dependent transitions. The temperatures are also labeled next to the lines in the corresponding colors to facilitate analysis.
    }
        \label{AVC}
    \end{figure}
    In Figs.\ref{ISO}(a) and (b), we observe that isolated spins exhibit continuous changes and peak as the system undergoes a primary phase transition of the second order. This phenomenon occurs as the temperature increases, leading to a gradual rise in the number of isolated spins, which consequently causes the system to fragment. However, once the system reaches a certain level of fragmentation, these isolated spins begin to aggregate into smaller clusters, thereby reducing the number of isolated spins, ultimately leading to the emergence of this peak. Thus, we define the peak of isolated spins as representing the position of the third-order independent transition, marked by black lines in Figs. \ref{ISO} (a) and (b). By applying this approach, we determined the positions of the third-order independent transitions for $D = 1.0$ and $D = 1.5$ to be $T = 1.207$ and $T = 0.968$, respectively. Using microcanonical inflection point analysis, the corresponding values were found to be $T = 1.3641$ and $T = 1.1139$. The temperature difference of about 0.2 between the third-order independent transitions obtained by the two methods can be attributed to differences in the application of canonical and microcanonical methods, along with the discrete nature of energy levels (with an interval of 1) in the microcanonical analysis. These two factors together contribute to the observed systematic error. 
    
    In Figs. \ref{ISO}(c) and (d), we observe that isolated spins exhibit a discontinuous jump after reaching a certain proportion when the system undergoes a first-order primary phase transition. In the insets of Figs. \ref{ISO} (c) and (d), magnified views show the regions where these discontinuities occur, with black and blue vertical lines indicating the positions where the isolated spins begin to exhibit jumps and where the rate of change reaches its maximum, respectively. We identify these positions as those of the third-order independent transition and the phase transition point, respectively. We define these as the positions of the third-order independent transition and the phase transition point, respectively. This jump phenomenon occurs because, as the crystal field parameter $D$ reaches a certain value, bringing the system closer to the phase transition, the fraction of spins in the 0 state rapidly increases, causing spins and clusters in the 0 state to dominate the system. In this scenario, the energy required to break clusters in the 0 state is lower than that required to break clusters in the -1 and 1 states. This is because the energy required for a spin to flip from the 0 state to either -1 or 1 is smaller than the energy required for a spin to flip from +1 to -1 (or from -1 to +1). As a result, even with minimal temperature variations, the number of isolated spins increases abruptly. At the point where the rate of increase reaches its peak, a phase transition occurs in the system. Therefore, we define the position at which the isolated spins undergo discontinuous jumps as the signal of the third-order independent transition, and the position where the change in isolated spins is most pronounced as the phase transition point. 
    When the primary phase transition is of the first order, we find that the temperatures of the third-order independent transition and the phase transition point are very close. However, as shown in the insets of Figs. \ref{ISO} (c) and (d), the isolated spins exhibit two distinct states with a considerable difference between them. Similar conclusions were reached in our microcanonical inflection point analysis.

    By comparing Tables \ref{TC} and \ref{third order transitions}, we observe that when the primary phase transition of the system is of the first order, the positions of the third-order independent transition and the phase transition point, as determined by microcanonical inflection point analysis, are closely aligned. In Figs. \ref{1_965}(a) and \ref{1_97}(a), the ``backbending'' phenomenon is observed in the first derivative of entropy, $\beta(E)$. By analyzing Figs. \ref{1_965}(b)-(c) and \ref{1_97}(b)-(c), we found that the third-order independent transition occurs in the ``backbending'' region at lower energy levels, whereas the phase transition point occurs in the ``backbending'' region at higher energy levels. Consequently, the temperatures of the third-order independent transition and the phase transition point are very close; however, there is a significant difference in their corresponding energy levels, which indicates that they are two distinct states. By comparing Figs. \ref{1_965}(a) and \ref{1_97}(a), it can be seen that as the value of $D$ increases, the ``backbending'' effect in $\beta(E)$ intensifies. As a result, the positions of the third-order independent transition and the phase transition point to become increasingly close to each other. This phenomenon is also observed in the Potts model, suggesting that it is characteristic of first-order phase transitions. The ``backbending'' phenomenon is a characteristic signal of phase coexistence in first-order phase transitions. When the primary phase transition of the system is of the first order, the maximum error between the positions of the third-order independent transition obtained by both methods is 0.002, considered negligible.

    In Fig. \ref{AVC}, we present the temperature-dependent variations of the average perimeter and its first derivative across different parameters. Our primary focus is to analyze the behavior of the first derivative. In the figure, red vertical lines mark the positions of the critical points, while purple lines indicate the locations of the identified third-order dependent transitions. Corresponding temperature values are annotated next to these lines in matching colors. When compared with the results obtained from microcanonical inflection point analysis, we find that the discrepancies between the two methods are approximately 0.03. Additionally, in Fig. \ref{AVC}(g), we observe a notable distinction, where a weak signal of a third-order dependent transition is detected and marked by a purple vertical line. Through careful analysis of this signal, we observed that the local minimum shifts upward as the system size increases. Consequently, we infer that, in the limit of infinite system size, this signal vanishes, aligning with the results from our microcanonical analysis. In Fig. \ref{AVC}(h), for $D = 1.97$, where the system's primary phase transition is of the first order, it is observed that this signal completely disappears. This phenomenon further implies that third-order dependent transition signals may not exist when the system's primary phase transition is of the first order.
    
    \begin{figure}
        \begin{center}
    	\epsfig{figure=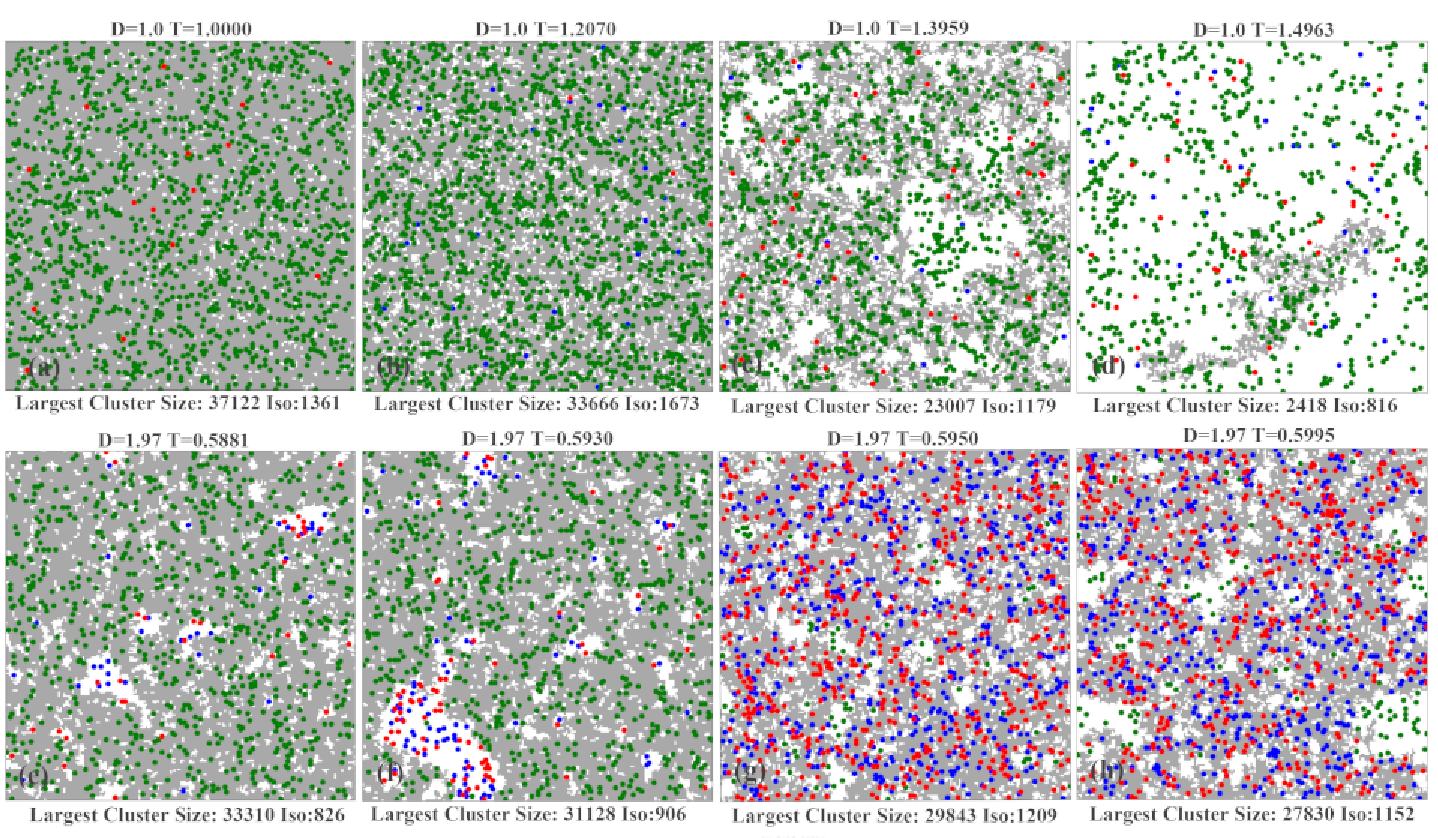,width=0.9\linewidth} \caption{(Distribution maps of isolated spins and the largest cluster).
    		This figure illustrates the spatial distribution of isolated spins and the largest cluster at different temperatures for various $D$ values, derived from simulations on a $200 \times 200$ system. The largest cluster is shown in gray, while all other clusters are displayed uniformly in white. Isolated spins are represented by dots, with blue, green, and red dots corresponding to the $-1$, $0$, and $+1$ spin states, respectively.
    		}
            \label{ISOG}
        \end{center}
    \end{figure}
    We thoroughly analyzed the distribution of isolated spins, the largest cluster, and the overall cluster structure to elucidate the characteristics of the third-order transition. Figures \ref{ISOG}(a)-(d) depict the distribution of isolated spins and the largest cluster when the primary phase transition of the system is of the second order ($D = 1.0$). In these figures, the largest cluster is represented in gray, while all other clusters are consistently shown in white. Isolated spins are shown as dots, with blue, green, and red dots representing isolated spins in the $-1$, $0$, and $+1$ spin states, respectively. The figures reveal that the number of isolated spins peaks near the third-order independent transition point. Beyond this point, the largest cluster becomes increasingly fragmented by isolated spins, which subsequently coalesce into smaller clusters, resulting in a decline in their number. Figures \ref{ISOG}(e)-(h) depict the distribution of isolated spins and the largest cluster when the primary phase transition of the system is of the first order ($D = 1.97$). We observe that isolated spins in the $0$ state are dominant before the third-order independent transition, while isolated spins in the $-1$ and $+1$ states emerge as predominant after the transition. These figures offer ample evidence supporting our explanation of the abrupt increase in the number of isolated spins as observed in Figures \ref{ISO}(c)-(d).
    
    \begin{figure}
        \begin{center}
		\epsfig{figure=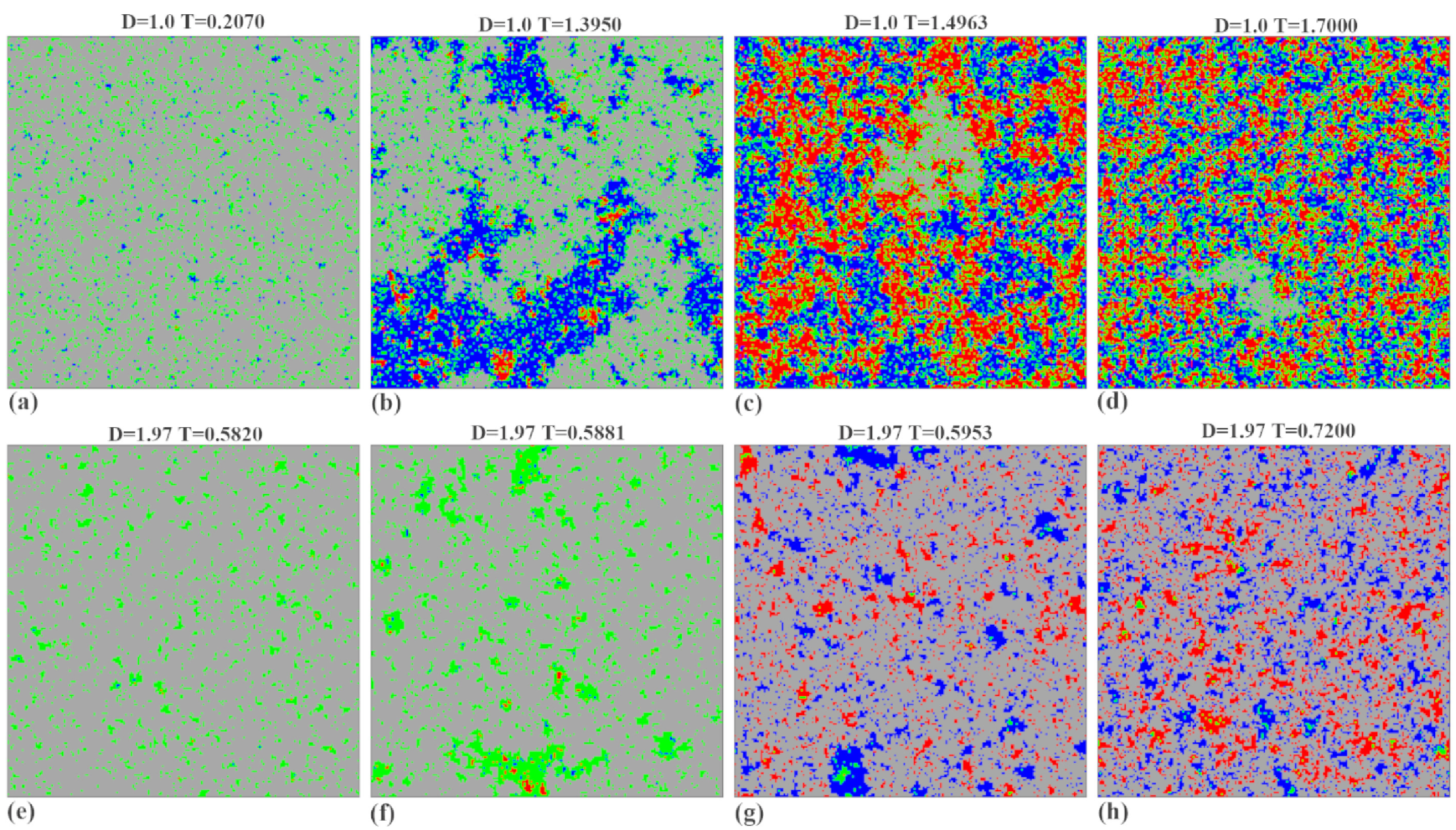,width=0.9\linewidth} \caption{(Cluster distribution map).
		This figure illustrates the spatial distribution of clusters at different temperatures for various $D$ values, based on simulations of a $200 \times 200$ system. The largest cluster is shown in gray, while clusters in the $-1$, $0$, and $+1$ spin states are depicted in blue, green, and red, respectively.
		}
            \label{cluster}
        \end{center}
    \end{figure}
    Figure \ref{cluster} illustrates the cluster distribution at various $D$ values and temperatures $T$. In the figure, the largest cluster is shown in gray, while clusters in the $-1$, $0$, and $+1$ states are depicted in blue, green, and red, respectively. Figures \ref{cluster}(a)-(d) display the cluster distribution when the primary phase transition of the system is of the second order ($D = 1.0$). As the temperature increases, the largest cluster gradually breaks down into smaller clusters that are still larger than typical small clusters, before further fragmenting into even smaller clusters. This process progresses continuously, manifesting macroscopically as a continuous phase transition, accompanied by the emergence of a third-order dependent transition. 
    Figures \ref{cluster}(e)-(f) illustrate the cluster distribution when the primary phase transition of the system is of the first order ($D = 1.97$). The largest cluster also gradually ``dissolves"; however, within the temperature range of our simulation, it persists throughout the system. However, near the phase transition point, as the temperature $T$ increases, large clusters undergo a sudden transition from the -1 state to the 0 state. This phenomenon is clearly analysed in Figures \ref{cluster}(g) and (h) and has been validated through Metropolis sampling, which confirms this behavior. In fact, it represents a phase transition from an ordered ferromagnetic state to an ordered paramagnetic state.
    Analysis of cluster changes during the cooling process reveals that smaller clusters gradually grow but fail to form significantly larger clusters. 
    We hypothesize that during cooling, smaller clusters continuously flip and may transition into states identical to the largest cluster, ultimately merging with it. This dynamic process inhibits the formation of larger cluster blocks, resulting in discontinuous changes in the order parameter. Therefore, when the primary phase transition of the system is of the first order, no third-order dependent transition signals are detected. Macroscopically, this phenomenon is marked by the discontinuity of the first-order phase transition. From these results, we find that the third-order transition is a phenomenon that emerges as the phase transition process.
    \begin{figure}
        \begin{center}
            \epsfig{figure=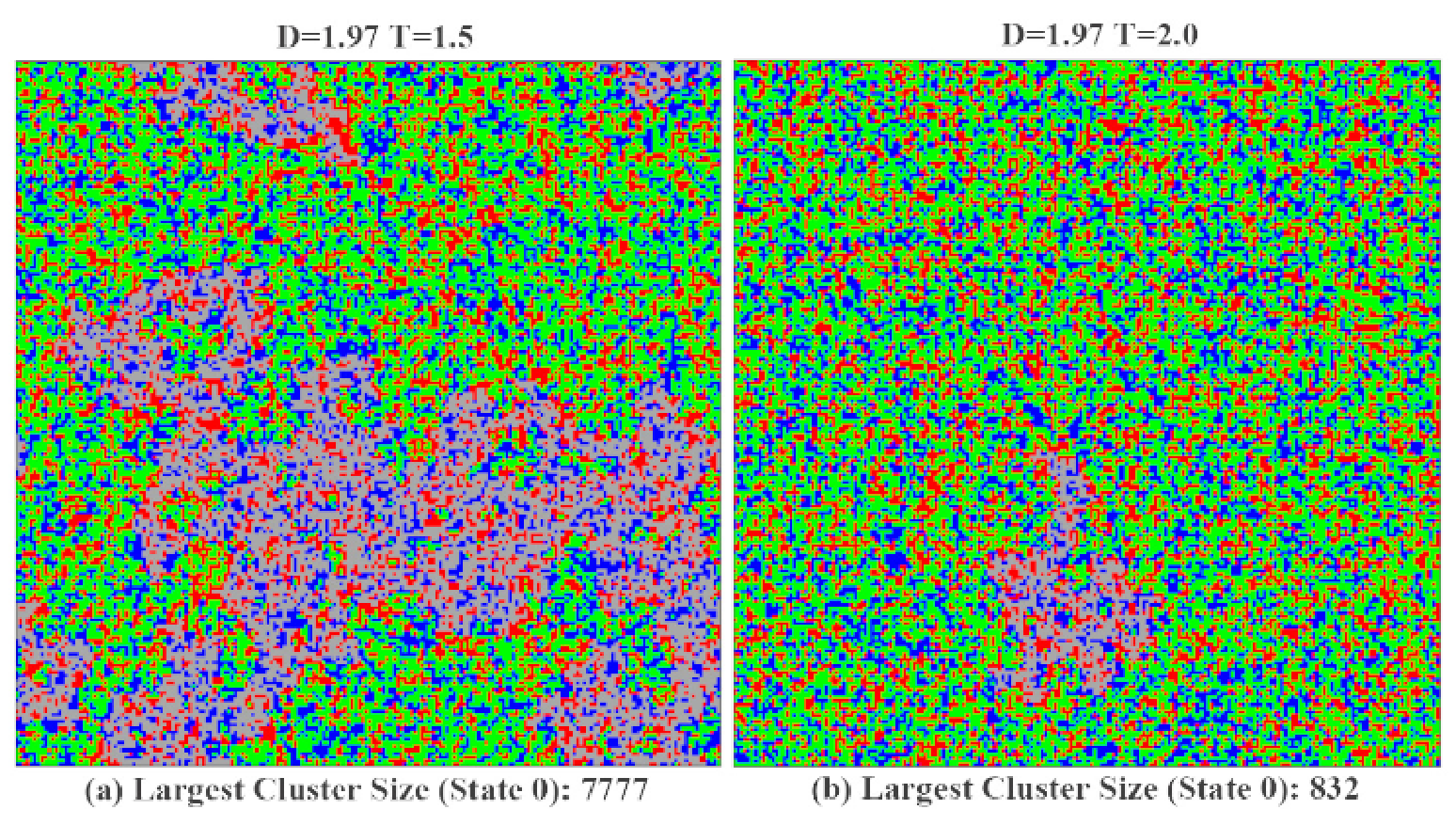,width=0.5\linewidth}
            \caption{(Cluster distribution before and after the fourth-order independent transition).
            This figure illustrates the spatial distribution of clusters at different temperatures for various $D$ values, based on simulations of a $200 \times 200$ system. The largest cluster in the 0 state is shown in gray, while clusters in the $-1$, $0$, and $+1$ spin states are depicted in blue, green, and red, respectively.
            }
            \label{1_97configuration}
        \end{center}
    \end{figure}
    
\begin{figure}
        \begin{center}
            \epsfig{figure=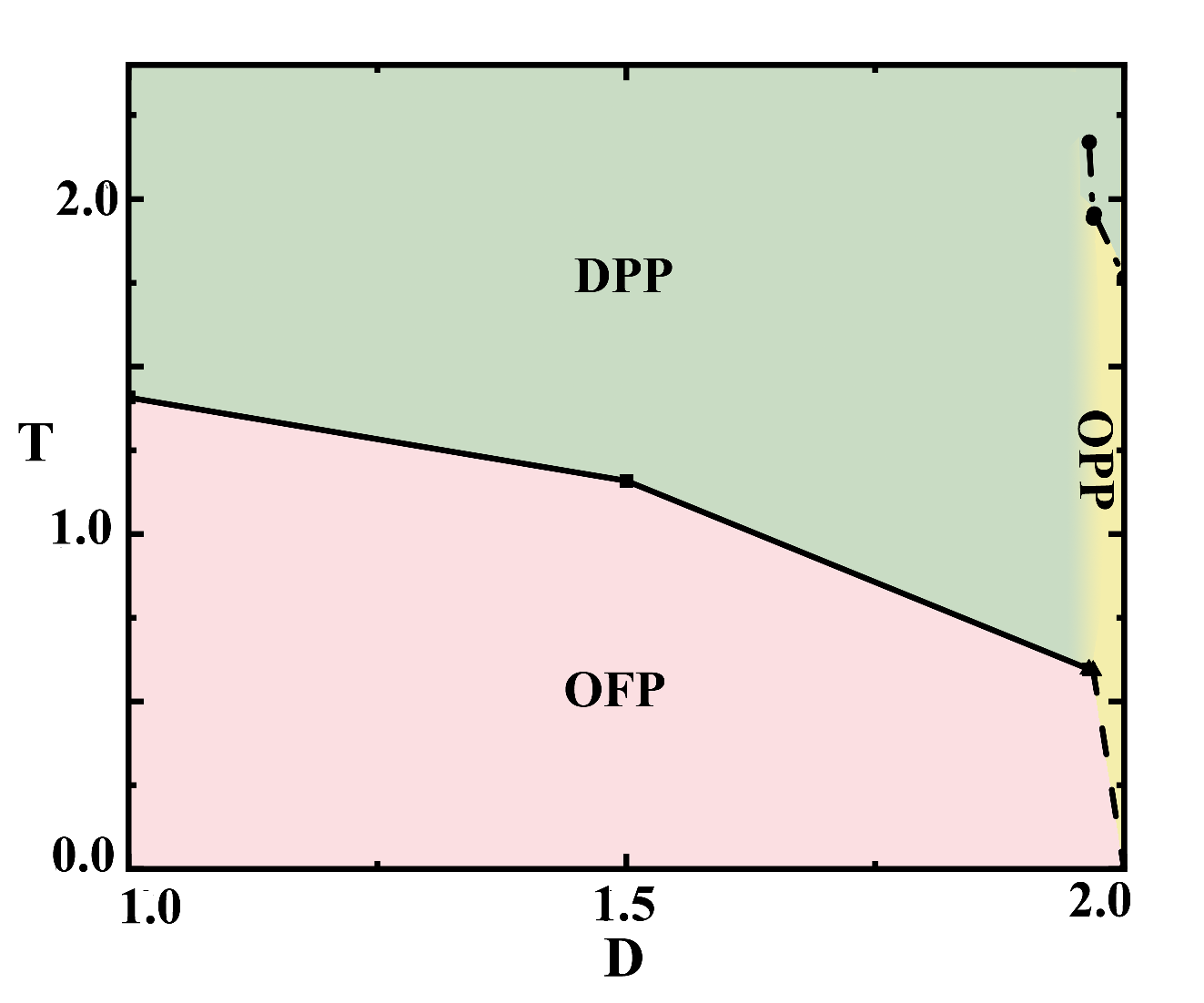,width=0.4\linewidth}
            \caption{(Blume-Capel model phase map).
            This diagram expands on the previous phase diagram of the Blume-Capel model. OFP denotes the ordered ferromagnetic phase, DPP denotes the disordered paramagnetic phase, and OPP denotes the ordered paramagnetic phase. Second-order phase transitions are indicated by solid lines, first-order transitions by dashed lines, and the boundary for the transition from the ordered paramagnetic phase to the disordered paramagnetic phase when \(D\) exceeds 1.965 is marked by a dotted line.
            }
            \label{phase}
        \end{center}
    \end{figure}

    As previously noted, when the primary phase transition of the system is of the first order, it shifts from an ordered ferromagnetic state to an ordered paramagnetic transitions. We hypothesize that, with a further increase in temperature, a transition from the ordered paramagnetic state to the disordered paramagnetic state may occur. To investigate this phenomenon, we performed a more detailed analysis of the high-temperature region using the microcanonical inflection point analysis method. A fourth-order independent transition was observed at $D = 1.965$, $1.969$, $1.97$, and $2.0$. In the inset of Figure \ref{1_97} (d), the position of the fourth-order independent transition is indicated by a purple vertical line. Furthermore, the corresponding inset in Figure \ref{1_97} (a) is provided to assist in extracting the relevant temperature information. 
    This result was further validated by Metropolis sampling. In Figure \ref{1_97configuration}, we present snapshots of the system before and after the fourth-order independent transition at $D = 1.97$. We observed that the temperature at which the ordered paramagnetic phase transitions to the disordered paramagnetic phase is very close to the temperature of the fourth-order independent transition. This indicates that the fourth-order independent transition acts as a marker for the onset of complete disorder in the high-temperature region and can function as a warning indicator in high-temperature systems.
    The data were organized and presented in the form of a phase diagram in Figure \ref{phase}. In the diagram, OFP denotes the ordered ferromagnetic phase, DPP denotes the disordered paramagnetic phase, and OPP denotes the ordered paramagnetic phase. Information regarding the phase transition points is obtained through microcanonical inflection point analysis. Near $D=1.9$, data for three specific points $D=1.965$, $1.969$, and $1.97$ detail the phase transition points. Second-order phase transitions are indicated by solid lines, first-order transitions by dashed lines, and the transition from the ordered paramagnetic phase to the disordered paramagnetic phase at high temperatures—when the primary phase transition is of the first order—is marked by a dotted line. When the primary phase transition of the system is of the first order, the system undergoes a transition from the ordered paramagnetic phase to the disordered paramagnetic phase through a fourth-order independent transition.
    
    Using two different methods, we identified signals of both third-order independent and dependent transitions within the second-order phase transition regime, while only third-order independent transition signals were detected in the first-order phase transition regime. Through microcanonical inflection point analysis and Metropolis sampling, we determined that when $D$ exceeds 1.965, the system transitions from the ordered ferromagnetic state to the ordered paramagnetic state, and eventually to the disordered paramagnetic state. The transition from the ordered paramagnetic state to the disordered paramagnetic state corresponds to fourth-order independent transition. We extend the phase diagram of the Blume-Capel model.
    
    \section{Summary}
    In this paper, we employ the Wang-Landau (WL) sampling method to obtain the density of states (DOS) for the Blume-Capel model, while using the Metropolis sampling method to simulate the spin configurations and extract information on isolated spins and the average cluster perimeter. We performed both canonical and microcanonical ensemble analyses on the data extracted from these two methods, successfully determining the positions of the critical and first-order phase transition points. The results from both sampling methods showed only minor differences, providing strong evidence for the validity and accuracy of our findings.

    Importantly, our study revealed two intriguing phenomena. One of these is that through microcanonical inflection point analysis, we observed that both third-order dependent and independent transitions coexist at the critical phase transition, while only third-order independent transitions were observed in the first-order phase transition. This observation was further verified through geometric analysis of the behavior of isolated spins and average cluster perimeters. Similar conclusions have been drawn in the Ising model \cite{sitarachu2022evidence} and the Potts model \cite{wang2024exploring}. Through the analysis of isolated spin distributions, largest cluster distributions, and the evolution of clusters, we found that third-order transitions may represent a phenomenon that accompanies the occurrence of phase transitions. Building on these findings, we further hypothesize that third-order dependent transitions vanish in the presence of strong first-order phase transitions. 
    Another finding is that in the high-temperature region of the first-order phase transition in the Blume-Capel model, we used microcanonical inflection point analysis to identify a fourth-order independent phase transition. Our Metropolis sampling analysis of the largest cluster of zero spins suggests that the temperature of the fourth-order independent transition aligns with the transition temperature from the ordered paramagnetic phase to the disordered paramagnetic phase. This discovery further enriches the existing phase diagram of the Blume-Capel model. These two discoveries offer significant theoretical support for our research on critical early-warning systems and provide valuable guidance for future studies on phase transitions.

    This method has provided us with more comprehensive information about the system, while the identified third-order transitions form a crucial basis for our early warning efforts. We intend to utilize this method for further exploration.

\section{Acknowledgments}
We would to thank Professor Youjin Deng for his valuable discussions. This work is supported by the National Natural Science Foundation of China No.12304257.

\bibliographystyle{elsarticle-num} 
\bibliography{cankaowengxian}

\begin{thebibliography}{10}
\expandafter\ifx\csname url\endcsname\relax
  \def\url#1{\texttt{#1}}\fi
\expandafter\ifx\csname urlprefix\endcsname\relax\def\urlprefix{URL }\fi
\expandafter\ifx\csname href\endcsname\relax
  \def\href#1#2{#2} \def\path#1{#1}\fi

\bibitem{gross2001microcanonical}
D.~H. Gross, Microcanonical thermodynamics: phase transitions in" small" systems, Vol.~66, World Scientific, 2001.

\bibitem{franzosi1999topological}
R.~Franzosi, L.~Casetti, L.~Spinelli, M.~Pettini, Topological aspects of geometrical signatures of phase transitions, Physical Review E 60~(5) (1999) R5009.

\bibitem{casetti2000geometric}
L.~Casetti, M.~Pettini, E.~Cohen, Geometric approach to hamiltonian dynamics and statistical mechanics, Physics Reports 337~(3) (2000) 237--341.

\bibitem{lenton2008tipping}
T.~M. Lenton, H.~Held, E.~Kriegler, J.~W. Hall, W.~Lucht, S.~Rahmstorf, H.~J. Schellnhuber, Tipping elements in the earth's climate system, Proceedings of the national Academy of Sciences 105~(6) (2008) 1786--1793.

\bibitem{fan2018climate}
J.~Fan, J.~Meng, Y.~Ashkenazy, S.~Havlin, H.~J. Schellnhuber, Climate network percolation reveals the expansion and weakening of the tropical component under global warming, Proceedings of the National Academy of Sciences 115~(52) (2018) E12128--E12134.

\bibitem{scheffer2003catastrophic}
M.~Scheffer, S.~R. Carpenter, Catastrophic regime shifts in ecosystems: linking theory to observation, Trends in ecology \& evolution 18~(12) (2003) 648--656.

\bibitem{arumugam2024early}
R.~Arumugam, F.~Guichard, F.~Lutscher, Early warning indicators capture catastrophic transitions driven by explicit rates of environmental change, Ecology 105~(4) (2024) e4240.

\bibitem{donangelo2010early}
R.~Donangelo, H.~Fort, V.~Dakos, M.~Scheffer, E.~H. Van~Nes, Early warnings for catastrophic shifts in ecosystems: Comparison between spatial and temporal indicators, International Journal of Bifurcation and Chaos 20~(02) (2010) 315--321.

\bibitem{brovkin2021past}
V.~Brovkin, E.~Brook, J.~W. Williams, S.~Bathiany, T.~M. Lenton, M.~Barton, R.~M. DeConto, J.~F. Donges, A.~Ganopolski, J.~McManus, et~al., Past abrupt changes, tipping points and cascading impacts in the earth system, Nature Geoscience 14~(8) (2021) 550--558.

\bibitem{dmitriev2023twitter}
A.~Dmitriev, A.~Lebedev, V.~Kornilov, V.~Dmitriev, Twitter self-organization to the edge of a phase transition: Discrete-time model and effective early warning signals in phase space, Complexity 2023~(1) (2023) 3315750.

\bibitem{dakos2010spatial}
V.~Dakos, E.~H. van Nes, R.~Donangelo, H.~Fort, M.~Scheffer, Spatial correlation as leading indicator of catastrophic shifts, Theoretical Ecology 3 (2010) 163--174.

\bibitem{rietkerk2021evasion}
M.~Rietkerk, R.~Bastiaansen, S.~Banerjee, J.~van~de Koppel, M.~Baudena, A.~Doelman, Evasion of tipping in complex systems through spatial pattern formation, Science 374~(6564) (2021) eabj0359.

\bibitem{bian2024early}
J.~Bian, Z.~Ma, C.~Wang, T.~Huang, C.~Zeng, Early warning for spatial ecological system: Fractal dimension and deep learning, Physica A: Statistical Mechanics and its Applications 633 (2024) 129401.

\bibitem{yan2023thermodynamic}
H.~Yan, F.~Zhang, J.~Wang, Thermodynamic and dynamical predictions for bifurcations and non-equilibrium phase transitions, Communications Physics 6~(1) (2023) 110.

\bibitem{bury2021deep}
T.~M. Bury, R.~Sujith, I.~Pavithran, M.~Scheffer, T.~M. Lenton, M.~Anand, C.~T. Bauch, Deep learning for early warning signals of tipping points, Proceedings of the National Academy of Sciences 118~(39) (2021) e2106140118.

\bibitem{junghans2006microcanonical}
C.~Junghans, M.~Bachmann, W.~Janke, Microcanonical analyses of peptide aggregation processes, Physical review letters 97~(21) (2006) 218103.

\bibitem{qi2018classification}
K.~Qi, M.~Bachmann, Classification of phase transitions by microcanonical inflection-point analysis, Physical review letters 120~(18) (2018) 180601.

\bibitem{sitarachu2022evidence}
K.~Sitarachu, M.~Bachmann, Evidence for additional third-order transitions in the two-dimensional ising model, Physical Review E 106~(1) (2022) 014134.

\bibitem{wang2024exploring}
F.~Wang, W.~Liu, J.~Ma, K.~Qi, Y.~Tang, Z.~Di, Exploring transitions in finite-size potts model: comparative analysis using wang--landau sampling and parallel tempering, Journal of Statistical Mechanics: Theory and Experiment 2024~(9) (2024) 093201.

\bibitem{liu2022pseudo}
W.~Liu, F.~Wang, P.~Sun, J.~Wang, Pseudo-phase transitions of ising and baxter--wu models in two-dimensional finite-size lattices, Journal of Statistical Mechanics: Theory and Experiment 2022~(9) (2022) 093206.

\bibitem{hong2024imaging}
J.~Hong, Y.~Tian, T.~Liang, X.~Liu, Y.~Song, D.~Guan, Z.~Yan, J.~Guo, B.~Tang, D.~Cao, et~al., Imaging surface structure and premelting of ice ih with atomic resolution, Nature (2024) 1--6.

\bibitem{blume1966theory}
M.~Blume, Theory of the first-order magnetic phase change in u o 2, Physical Review 141~(2) (1966) 517.

\bibitem{capel1966possibility}
H.~Capel, On the possibility of first-order phase transitions in ising systems of triplet ions with zero-field splitting, Physica 32~(5) (1966) 966--988.

\bibitem{saul1974tricritical}
D.~Saul, M.~Wortis, D.~Stauffer, Tricritical behavior of the blume-capel model, Physical Review B 9~(11) (1974) 4964.

\bibitem{gunaratnam2024existence}
T.~S. Gunaratnam, D.~Krachun, C.~Panagiotis, Existence of a tricritical point for the blume--capel model on $\mathbb{Z}d$, Probability and Mathematical Physics 5~(3) (2024) 785--845.

\bibitem{kwak2015first}
W.~Kwak, J.~Jeong, J.~Lee, D.-H. Kim, First-order phase transition and tricritical scaling behavior of the blume-capel model: A wang-landau sampling approach, Physical Review E 92~(2) (2015) 022134.

\bibitem{goldman1973tricritical}
A.~Goldman, Tricritical points of thin superconducting films, Physical Review Letters 30~(21) (1973) 1038.

\bibitem{moueddene2024critical}
L.~Moueddene, N.~G. Fytas, Y.~Holovatch, R.~Kenna, B.~Berche, Critical and tricritical singularities from small-scale monte carlo simulations: the blume--capel model in two dimensions, Journal of Statistical Mechanics: Theory and Experiment 2024~(2) (2024) 023206.

\bibitem{azhari2020tricritical}
M.~Azhari, U.~Yu, Tricritical point in the mixed-spin blume-capel model on three-dimensional lattices: Metropolis and wang-landau sampling approaches, Physical Review E 102~(4) (2020) 042113.

\bibitem{blote2019revisiting}
H.~W. Bl{\"o}te, Y.~Deng, Revisiting the field-driven edge transition of the tricritical two-dimensional blume-capel model, Physical Review E 99~(6) (2019) 062133.

\bibitem{mandal2016geometrical}
I.~Mandal, S.~Inglis, R.~G. Melko, Geometrical mutual information at the tricritical point of the two-dimensional blume--capel model, Journal of Statistical Mechanics: Theory and Experiment 2016~(7) (2016) 073105.

\bibitem{antenucci2014critical}
F.~Antenucci, A.~Crisanti, L.~Leuzzi, Critical study of hierarchical lattice renormalization group in magnetic ordered and quenched disordered systems: Ising and blume--emery--griffiths models, Journal of Statistical Physics 155 (2014) 909--931.

\bibitem{mahan1978blume}
G.~Mahan, S.~Girvin, Blume-capel model for plane-triangular and fcc lattices, Physical Review B 17~(11) (1978) 4411.

\bibitem{clusel2008alternative}
M.~Clusel, J.-Y. Fortin, V.~N. Plechko, Alternative description of the 2d blume--capel model using grassmann algebra, Journal of Physics A: Mathematical and Theoretical 41~(40) (2008) 405004.

\bibitem{silva2006wang}
C.~Silva, A.~Caparica, J.~Plascak, Wang-landau monte carlo simulation of the blume-capel model, Physical Review E—Statistical, Nonlinear, and Soft Matter Physics 73~(3) (2006) 036702.

\bibitem{dias2011site}
D.~Dias, J.~Plascak, Site-diluted blume--capel model for the fe--al disordered alloys, Physics Letters A 375~(21) (2011) 2089--2093.

\bibitem{pena2009blume}
D.~Pe{\~n}a~Lara, G.~P{\'e}rez~Alc{\'a}zar, L.~E. Zamora, J.~Plascak, Blume-capel model for (fe 0.65 ni 0.35) 1- x mn x and fe p al q mn x alloys, Physical Review B—Condensed Matter and Materials Physics 80~(1) (2009) 014427.

\bibitem{blatter1985reversible}
A.~Blatter, M.~Von~Allmen, Reversible amorphization in laser-quenched titanium alloys, Physical review letters 54~(19) (1985) 2103.

\bibitem{sinkler1997neutron}
W.~Sinkler, C.~Michaelsen, R.~Bormann, D.~Spilsbury, N.~Cowlam, Neutron-diffraction investigation of structural changes during inverse melting of ti 45 cr 55 s, Physical Review B 55~(5) (1997) 2874.

\bibitem{mazzitello2015far}
K.~I. Mazzitello, J.~Candia, E.~V. Albano, Far-from-equilibrium growth of magnetic thin films with blume-capel impurities, Physical Review E 91~(4) (2015) 042118.

\bibitem{kaufman2024social}
M.~Kaufman, S.~Kaufman, H.~T. Diep, Social depolarization: Blume--capel model, Physics 6~(1) (2024) 138--147.

\bibitem{diep2024monte}
H.~T. Diep, M.~Kaufman, S.~Kaufman, Monte carlo study of agent-based blume-capel model for political depolarization, in: EPJ Web of Conferences, Vol. 300, EDP Sciences, 2024, p. 01005.

\bibitem{mozolenko2024blume}
V.~Mozolenko, L.~Shchur, Blume-capel model analysis with a microcanonical population annealing method, Physical Review E 109~(4) (2024) 045306.

\bibitem{cirillo2024homogeneous}
E.~N. Cirillo, V.~Jacquier, C.~Spitoni, Homogeneous and heterogeneous nucleation in the three-state blume--capel model, Physica D: Nonlinear Phenomena 461 (2024) 134125.

\bibitem{akin2024investigation}
H.~Ak{\i}n, Investigation of thermodynamic properties of mixed-spin (2, 1/2) ising and blum--capel models on a cayley tree, Chaos, Solitons \& Fractals 184 (2024) 114980.

\bibitem{ovchinnikov2024influence}
A.~Ovchinnikov, I.~Bostrem, V.~E. Sinitsyn, N.~Nosova, N.~Baranov, Influence of the type of intercalation on spin-glass formation in the fe-doped tas 2 (se 2) polytype family, Physical Review B 109~(5) (2024) 054403.

\bibitem{bechinger2016active}
C.~Bechinger, R.~Di~Leonardo, H.~L{\"o}wen, C.~Reichhardt, G.~Volpe, G.~Volpe, Active particles in complex and crowded environments, Reviews of modern physics 88~(4) (2016) 045006.

\bibitem{copeland2009bacterial}
M.~F. Copeland, D.~B. Weibel, Bacterial swarming: a model system for studying dynamic self-assembly, Soft matter 5~(6) (2009) 1174--1187.

\bibitem{qi2022emergence}
K.~Qi, E.~Westphal, G.~Gompper, R.~G. Winkler, Emergence of active turbulence in microswimmer suspensions due to active hydrodynamic stress and volume exclusion, Communications Physics 5~(1) (2022) 49.

\bibitem{wang2001efficient}
F.~Wang, D.~P. Landau, Efficient, multiple-range random walk algorithm to calculate the density of states, Physical review letters 86~(10) (2001) 2050.

\bibitem{landau2004new}
D.~Landau, S.-H. Tsai, M.~Exler, A new approach to monte carlo simulations in statistical physics: Wang-landau sampling, American Journal of Physics 72~(10) (2004) 1294--1302.

\bibitem{li2007numerical}
Y.~W. Li, T.~W{\"u}st, D.~P. Landau, H.~Lin, Numerical integration using wang--landau sampling, Computer physics communications 177~(6) (2007) 524--529.

\bibitem{wust2011unraveling}
T.~W{\"u}st, Y.~W. Li, D.~P. Landau, Unraveling the beautiful complexity of simple lattice model polymers and proteins using wang-landau sampling, Journal of Statistical Physics 144~(3) (2011) 638--651.

\bibitem{vogel2013generic}
T.~Vogel, Y.~W. Li, T.~W{\"u}st, D.~P. Landau, Generic, hierarchical framework for massively parallel wang-landau sampling, Physical review letters 110~(21) (2013) 210603.

\bibitem{vogel2014scalable}
T.~Vogel, Y.~W. Li, T.~W{\"u}st, D.~P. Landau, Scalable replica-exchange framework for wang-landau sampling, Physical Review E 90~(2) (2014) 023302.

\bibitem{ferreira2018wang}
L.~Ferreira, L.~Jorge, S.~Le{\~a}o, A.~Caparica, Wang--landau sampling: Saving cpu time, Journal of Computational Physics 358 (2018) 130--134.

\bibitem{bachmann2014thermodynamics}
M.~Bachmann, Thermodynamics and statistical mechanics of macromolecular systems, Cambridge University Press, 2014.

\bibitem{metropolis1953equation}
N.~Metropolis, A.~W. Rosenbluth, M.~N. Rosenbluth, A.~H. Teller, E.~Teller, Equation of state calculations by fast computing machines, The journal of chemical physics 21~(6) (1953) 1087--1092.

\bibitem{fernandes2015critical}
F.~Fernandes, D.~F. de~Albuquerque, F.~Lima, J.~Plascak, Critical behavior of the spin-1 blume-capel model on two-dimensional voronoi-delaunay random lattices, Physical Review E 92~(2) (2015) 022144.

\bibitem{beale1986finite}
P.~D. Beale, Finite-size scaling study of the two-dimensional blume-capel model, Physical Review B 33~(3) (1986) 1717.

\end{thebibliography}

\end{document}